\documentclass{aa}
\usepackage{graphics}

\begin{document}

   \title{Turbulent viscosity in clumpy accretion disks II\\
        supernova driven turbulence in the Galaxy}

   \titlerunning{Turbulent viscosity in clumpy accretion disks II}

   \author{B.~Vollmer \and  T.~Beckert}

   \offprints{B.~Vollmer, e-mail: bvollmer@mpifr-bonn.mpg.de}

   \institute{Max-Planck-Institut f\"ur Radioastronomie, Auf dem
              H\"ugel 69, 53121 Bonn, Germany.}

   \date{Received / Accepted}

\abstract{
An analytical model for a turbulent clumpy gas disk is presented where
turbulence is maintained by the energy input due to supernovae.
Expressions for the disk parameters, global filling factors,
molecular fractions, and star formation rates are given as functions
of the Toomre parameter $Q$, the ratio between the cloud size
and the turbulent driving length scale $\delta$,
the mass accretion rate within the disk $\dot{M}$,
the constant of molecule formation $\alpha$,
the disk radius, the angular velocity, and its radial derivative.
Two different cases are investigated: a dominating stellar disk and
a self-gravitating gas disk in $z$ direction.
The turbulent driving wavelength is determined in a first
approach by energy flux conservation, i.e. the supernovae energy input
is transported by turbulence to smaller scales where it is dissipated.
The results are compared to those of a fully gravitational model.
For $Q=1$ and $\delta=1$ both models are consistent with each other.
In a second approach the driving length scale is directly determined by the
size of supernovae remnants. Both models are applied to the Galaxy
and can reproduce its integrated and local gas properties.
The influence of thermal and magnetic pressure on the disk structure
is investigated. We infer $Q \sim 1$ and
$\dot{M} \sim 0.05 - 0.1$~M$_{\odot}$yr$^{-1}$ for the Galaxy.
 }

\maketitle

\section{Introduction}

The interstellar medium (ISM) is of turbulent nature. Its structure
is usually described as hierarchical (Scalo 1985) over length scales of several
magnitudes up to $\sim$100~pc. The neutral phase of the
ISM is not uniform but of fractal nature (Elmegreen \& Falgarone 1996).

In a widely accepted picture (see, e.g. Kulkarni \& Heiles 1988,
Spitzer 1990, McKee 1995), the ISM consists of 5 different phases
that are listed in Table~\ref{tab:tab1}.
\begin{table*}
\begin{tabular}{l|ccccccc}
 & $T$\ ({\rm K}) & $n$\ ({\rm cm}$^{-3}$) & $<n>$\ ({\rm cm}$^{-3}$) & $v_{\rm rms}$\ ({\rm km\,s$^{-1}$}) & $\Phi_{\rm V}$ & $H$\ ({\rm pc}) & M/M$_{\rm tot}$ ($\%$)\\
\hline
{\rm hot\ ionized} & $\sim 10^{6}$ & .. & 0.002 & .. & 0.5 & 3000 & 4 \\
{\rm warm\ ionized} & $\sim 8000$ & 0.3--10 & 0.025 & $\sim$10 & 0.2 & 900 & 14 \\
{\rm warm\ neutral} & $\sim 8000$ & 0.1--10 & 0.1 & $\sim$10 & 0.3 & 400 & 31 \\
{\rm cold\ neutral} & $\sim 100$ & 10--1000 & 0.3 & $\sim$6 & 0.02 & 140 & 25 \\
{\rm molecular} & $\sim 10$ & $>100$ & 0.6 & $\sim$6 & 0.001 & 70 & 26 \\
\end{tabular}
\label{tab:tab1}
\caption{The properties of the different phases of the ISM ($T$: temperature, $n$: density,
$<n>$:mean density, $v$: dispersion velocity, $\Phi_{\rm V}$: volume filling factor,
$H$: scale height, $M$: percentage of the total gas mass (from Boulares \& Cox 1990).}
\end{table*}
About 80\% of the total gas mass is neutral and 50\% is in form of clouds or
filaments. The scale height of the cold material is about 4 times smaller
than that of the warm gas.
The hot medium is heated and ionized by direct thermal energy input due to
supernova (SN) explosions. The warm medium is heated and ionized by SN remnants,
stellar winds and radiation.
The main energy sources driving interstellar
turbulence and thus causing the multiphase structure of the ISM are
(i) SN explosions (see, e.g., Ruzmaikin et al. 1988) or (ii) the galaxy's
gravitational potential and local gravitational instabilities (Wada et al. 2002).

Several attempts have been made to include SN explosions into
hydrodynamical simulations:

Rosen \& Bregman (1995) computed two-dimensional simulations with two
cospatial fluids representing stars and gas. Star formation, stellar mass
loss, stellar winds and SN were included. These simulations created a
three-phase medium with filaments of dense, cold and warm gas surrounding
bubbles of hot gas.

Korpi et al. (1999) simulated
the local ISM heated and stirred by SNe using a three-dimensional,
non-ideal MHD model in a box of $\sim$1~kpc. They used a
prescription for the SN rate deduced from observations. These simulations showed a
stationary multicomponent structure of the gas in a state of
developed turbulence. They found a turbulent cell size of 60~pc
in the warm phase.

Gazol-Patino \& Passot (1999) studied the
role of superbubbles in the evolution of the ISM at the kiloparsec scale.
The model incorporated the fully compressible MHD equations,
including parameterized cooling and heating, the Coriolis force,
shear, self-gravitation, and energy input by SNe in 2.5 dimensions.
The SN rate was determined by the gas flow itself.
Their simulations also lead to a state of highly compressible turbulence.
The inclusion of SNe into the model gave rise to superbubbles whose presence
leads to a more intermittent cycle of star formation.

Wada \& Norman (2001) made high-resolution two-dimensional
hydrodynamic simulations of the ISM including star formation and
energy input by SNe. In their simulations a globally stable
multiphase ISM formed. They found an energy spectrum of
$E(k) \propto k^{-2}$ and a turbulent length scale of 20-100~pc.

The influence of SN explosions on the ISM was studied by de
Avillez \& Mac Low (2001) with a three-dimensional hydrodynamical code
using adaptive mesh refinement. SN explosions were set up at a rate
comparable to observations. They showed that hot gas that is not evacuated
through chimneys expands into the cooler gas of the thick disk building
mushroom-shaped structures.

Even though it is clear that SN explosions represent the dominant energy input
in galaxies with a normal star formation rate,
all these simulations show how SNe can effectively drive ISM turbulence in detail.

In order to form massive stars which have a SN explosion at the
end of their lifetime, self-gravitation is the starting point.
Self-gravitation drives star formation and subsequently SN explosions.

In a previous paper (Vollmer \& Beckert 2002) we have elaborated
a model of an equilibrium disk where turbulence
is generated by instabilities involving self-gravitation and
maintained by the energy input from differential rotation and
mass transfer. The model only involves gravity.
Its application to the Galaxy showed good agreement with observations.

In this article we extend our previous model by including SNe as
an additional energy input to drive ISM turbulence.
We give the basic picture in Sect.~\ref{sec:basic}. The equations
are presented in Sect.~\ref{sec:equations}. Two different models with
different vertical pressure equilibria are analyzed: (i) a model
of a self-gravitating gas disk in $z$ direction (SGZ model) and (ii) a model
with a dominating stellar disk mass (DSD model). We introduce an energy
flux conservation equation that relates the energy flux transported by turbulence
to the energy input due to SN explosions. In Sect.~\ref{sec:results} we give
analytical expressions for the disk properties and show results for
characteristic values of the input parameters. The SGZ model is compared
in Sect.~\ref{sec:fullgrav} to
the fully gravitational model of Vollmer \& Beckert (2002).
In Sect.~\ref{sec:snr} the energy flux conservation is replaced by
an equation relating the turbulent driving length scale directly to
the size of a SN remnant (SNR model). We apply the SGZ and SNR model to
the Galaxy in Sect.~\ref{sec:galaxy}. In Sect.~\ref{sec:realpot} a
realistic gravitational potential for the galaxy is introduced and the
inclusion of magnetic and thermal pressure into the models is discussed.
Sect.~\ref{sec:starformation} investigates disks with $Q > 1$ and $Q < 1$.
The summary and conclusions are given in Sect.~\ref{sec:summary}.

\section{The basic picture \label{sec:basic}}

We consider the warm, cold, and molecular phases of the ISM as one
gas that can undergo phase changes according to internal and external
conditions. Internal conditions are pressure, metalicity, molecule formation,
and turbulent time scales; external conditions are stellar radiation field
(UV, X-rays, cosmic rays), stellar winds, and SN explosions with
subsequent shock formation.
In this approach we neglect the thermal balance due to
radiative heating and cooling of the ISM.

The ISM is assumed to be turbulent. Turbulence is driven by
SNe energy input at a characteristic length scale $l_{\rm driv}$.
This length scale might be identified as the characteristic length scale
of a SN bubble or alternatively that of the interaction of SN bubbles.
The disk scale height is determined unambiguously
by the turbulent pressure $p_{\rm turb}=\rho v_{\rm turb}^{2}$,
where $\rho$ is the average density and $v_{\rm turb}$
the turbulent velocity in the disk.
We thus neglect thermal, cosmic ray, and magnetic pressure.
The interaction of SN bubbles leads to viscous transport of angular momentum.
We assume that the energy input due to SNe is dissipated by viscous heating.

Thus, we consider a gaseous turbulent accretion disk in a given gravitational
potential $\Phi$ which gives rise to an angular velocity
$\Omega=\sqrt{R^{-1}\frac{{\rm d}\Phi}{{\rm d}R}}$.
The Toomre parameter (Toomre 1964) is treated as a constant but free parameter
for the whole disk
\begin{equation}
Q=\frac{v_{\rm turb}\,\kappa}{\pi\,G\,\Sigma}\ ,
\end{equation}
with the restriction $Q \geq 1$, where $G$ is the gravitational constant,
$\Sigma$ the average gas surface density, and $\kappa$ the local
epicyclic frequency.

\section{The equations \label{sec:equations}}

\subsection{The volume filling factor \label{sec:vff}}

We compare the crossing time of a turbulent cloud to the
gravitational free fall time in order to derive an expression for the
volume filling factor $\phi_{\rm V}$.
The characteristic turbulent time scale of clouds whose size is a factor
$\delta^{-\frac{3}{1+D}}$ smaller than
the driving length scale $l_{\rm cl}=\delta^{-\frac{3}{1+D}}\,l_{\rm driv}$ is
\begin{equation}
t_{\rm l}=\delta^{-\frac{2}{3}-\frac{3-D}{3}}
\,l_{\rm driv}/v_{\rm turb}\ ,
\end{equation}
where $D$ is the fractal dimension (see, e.g., Frisch 1995).
The local gravitational free fall time is given by
\begin{equation}
t^{\rm l}_{\rm ff}=\sqrt{\frac{3\pi}{32G\rho_{\rm cl}}}\ ,
\label{eq:localff}
\end{equation}
where $\rho_{\rm cl}$ is the density of a single cloud, which is related
to the overall disk density $\rho$ by the volume filling factor $\phi_{\rm V}$:
$\rho_{\rm cl}=\phi_{\rm V}^{-1}\rho$.
The clouds become self-gravitating for $t_{\rm l}=t_{\rm ff}^{\rm l}$.
We assume $D=2$ for a compressible, selfgravitating fluid, which is
close to the findings of Elmegreen \& Falgarone (1996).

\subsection{The viscosity prescription \label{sec:vispres}}

In general the viscosity $\nu$ is defined as the product of the
characteristic velocity $v$ and the characteristic length scale $l$
of the system: $\nu = v l$. We assume that turbulence
becomes intermittent due to self-gravitation of cold gas clouds.
There are two simple frameworks for understanding intermittency:
(i) based on velocity (ii) based on dissipation.

Within the framework of intermittency based on dissipation (Frisch 1995),
the local spatial average of the energy dissipation rate per mass unit $\epsilon$,
which is constant over all length scales in the case of a Kolmogorov-like
turbulence, depends explicitly on the length scale
\begin{equation}
\frac{\epsilon_{l}}{v_{\rm turb}^{3}/l_{\rm driv}} \sim
\left(\frac{l}{l_{\rm driv}}\right)^{3-D}
\end{equation}
where $D$ is the fractal dimension ($D$=3 for a Kolmogorov-like turbulence),
$l_{\rm driv}$ is the driving length scale.
This can be achieved if there is an energy sink in the turbulent cascade.
We identify the self-gravitation of cold
gas clouds as this energy sink, i.e. during the contraction of a cloud
kinematic energy is transformed into heat and radiated away so that the cloud
can further shrink. Thus self-gravity gives rise to intermittency in the ISM turbulence.

Within the framework of intermittency based on velocity, intermittency can be
understood as a decreasing volume filling factor of turbulent eddies with
decreasing length scale. This can also be achieved by self-gravitation,
because self-gravitating clouds are smaller than diffuse clouds
of the same mass.

In the case of intermittent turbulence the turbulent viscosity
$\nu = v_{\rm turb} l_{\rm driv}$ overestimates the true viscosity and has to be
decreased by a factor $\gamma$:
\begin{equation}
\nu = \gamma v_{\rm turb} l_{\rm driv}\ .
\label{eq:gammavisc}
\end{equation}
This factor is related to self-gravitation, i.e.
the free fall time of the disk in $z$ direction.
Based on these arguments we make the hypothesis
that the characteristic velocity is determined by the disk height
$H$ and the averaged local free fall time $t^{\rm H}_{\rm ff}$:
$v = H/t^{\rm H}_{\rm ff}$. The appropriate length scale is the turbulent
driving length scale $l_{\rm driv}$.

Thus we obtain:
\begin{equation}
\nu = \frac{H}{t^{\rm H}_{\rm ff}} l_{\rm driv}\ .
\label{eq:nuu}
\end{equation}
The free fall time is given by $t^{\rm H}_{\rm ff}=\sqrt{(3\pi)/(32 G \rho)}$,
where $\rho$ is the averaged midplane density of the disk.
Using the local free fall time (Eq.~(\ref{eq:localff})) with $\rho_{\rm cl}=
\phi_{\rm V}^{-1} \rho$, where $\phi_{\rm V}$ is the volume filling factor of
self-gravitating clouds (Sect.~\ref{sec:vff}).
The characteristic velocity can then be written as
\begin{equation}
v = \frac{H}{t^{\rm H}_{\rm ff}}
 = \frac{H}{l_{\rm driv}} \sqrt{\phi_{\rm V}} \delta v_{\rm turb}\ .
\end{equation}
Inserting this velocity into Eq.~(\ref{eq:nuu}) gives
\begin{equation}
\nu = H \sqrt{\phi_{\rm V}} \delta v_{\rm turb}\ .
\label{eq:visc}
\end{equation}
Thus, the factor $\gamma$ that is due to intermittency is
\begin{equation}
\gamma = \frac{H}{l_{\rm driv}} \sqrt{\phi_{\rm V}} \delta \ .
\label{eq:intfactor}
\end{equation}
We will call it the viscosity intermittence factor.

\subsection{Angular momentum equation}

The viscosity prescription given above is essential for mass accretion and
the transport of angular momentum.
In a steady state accretion disk, the mass accretion rate is
\begin{equation}
\dot{M}=2\pi R\,\Sigma\,(-v_{\rm rad})\ ,
\end{equation}
where $v_{\rm rad}$ is the radial velocity.
The angular momentum equation can be integrated giving
\begin{equation}
\nu \Sigma = -\frac{\dot{M}}{2\pi R} \Omega \big(\frac{\partial \Omega}
{\partial R}\big)^{-1}\ .
\label{eq:ame}
\end{equation}
Furthermore, we use
\begin{equation}
\Sigma = \rho\,H
\label{eq:srh}
\end{equation}
for the surface density of the disk, where $\rho$ is the average mass
density at the midplane.

\subsection{The vertical pressure equilibrium}

We assume that
the only pressure which counterbalances gravitation in the
vertical direction is the turbulent pressure
$p_{\rm turb}=\rho v_{\rm turb}^{2}$. We distinguish two cases
for the gravitational force density in the vertical $z$ direction:
\begin{enumerate}
\item
$\rho \ll \rho_{*}$ and $M_{\rm d}(R) \ll M_{*}(R)$, where $\rho_{*}/M_{*}(R)$
is the stellar central density/disk mass
within a radius $R$ (dominating stellar disk mass);
\item
$\rho \geq \rho_{*}$ and $0.5\,(H/R)\,M(R) < M_{\rm d}(R) < M(R)$,
where $M(R)$ is the total mass enclosed within a radius $R$
(self-gravitating gas disk in $z$ direction).
\end{enumerate}
In the following we call the model of a dominating stellar disk mass
{\it DSD model} and the model of a self-gravitating gas disk in $z$
direction {\it SGZ model}.
The hydrostatic equilibrium in the vertical direction implies that the
gravitational force is balanced by the turbulent pressure.

For the two cases the hydrostatic equilibrium has the
following forms:
\begin{enumerate}
\item
\begin{equation}
p_{\rm turb}=\Sigma \partial \Phi / \partial z\ ,
\label{eq:pp1}
\end{equation}
where $\Phi$ is the gravitational
potential of the disk;
\item
\begin{equation}
p_{\rm turb}=\pi G \Sigma^{2}
\label{eq:pp2}
\end{equation}
 (Paczy\'{n}ski 1978).
\end{enumerate}

\subsection{Global gravitational stability in $z$ direction}

The basic principles underlying the gravitational instability
of a thin rotating disk can be found in Toomre (1964).
A gaseous disk is locally stable to axisymmetric perturbations, if
\begin{equation}
Q=\frac{v_{\rm turb}\,\kappa}{\pi G\,\Sigma} > 1\ ,
\end{equation}
where $\kappa=\sqrt{R\frac{{\rm d}\Omega^{2}}{{\rm d}R}+4\Omega^{2}}$
is the epicylcic frequency.
Since in general $\Omega \leq \kappa \leq 2\Omega$, we will use the
following equation:
\begin{equation}
Q \simeq \frac{v_{\rm turb}\,\Omega}{\pi\,G\,\Sigma}\ .
\label{eq:tq}
\end{equation}
Multiplying the numerator and the denominator of the right hand side
by $R^{2}$ gives
\begin{equation}
Q=\frac{v_{\rm turb}}{v_{\rm rot}}\frac{M_{\rm tot}}{M_{\rm gas}}\ ,
\label{eq:tqint}
\end{equation}
where $M_{\rm tot/gas}$ is the total enclosed mass and the total enclosed
gas mass at radius $R$.
Thus for a given velocity dispersion $Q^{-1}$ is proportional to the ratio
of gas mass to total mass.

\subsection{The star formation rate \label{sec:starform}}

It is generally accepted that the star formation rate is proportional to the mean
density of the disk and the inverse of
the characteristic time scale for the cloud collapse, i.e.
the non-averaged local free fall time $t^{\rm l}_{\rm ff}$ (see Sect.~\ref{sec:vff}):
\begin{equation}
\dot{\rho}_{*} \propto \frac{\rho}{t^{\rm l}_{\rm ff}}\ .
\label{eq:rhostar}
\end{equation}
Since $t^{\rm l}_{\rm ff} \propto \rho^{-\frac{1}{2}}$ (see Eq.~(\ref{eq:localff}))
this corresponds
to a Schmidt law of the form $\dot{\rho}_{*} \propto \rho^{\frac{3}{2}}$.
The factor of proportionality is given by the probability to find a
self-gravitating cloud, i.e. the volume filling factor $\phi_{\rm V}$.
Thus, the star formation rate is given by
\begin{equation}
\dot{\rho}_{*} = \phi_{\rm V} \frac{\rho}{t^{\rm l}_{\rm ff}}=
\sqrt{\phi_{\rm V}} \frac{\rho}{t_{\rm ff}^{\rm H}}\ .
\end{equation}
Furthermore, we assume that stars are only born in the midplane of the disk
in regions that have the size of the turbulent driving length scale $l_{\rm driv}$,
because the clouds can collapse only within the turbulent
time scale $t_{\rm turb}=l_{\rm driv}/v_{\rm turb}$. We thus obtain
\begin{equation}
\dot{\Sigma}_{*}=\dot{\rho}_{*}\,l_{\rm driv}
\label{eq:starform}
\end{equation}
for the mass surface density turned into stars.
The volume filling factor is defined such that $t_{\rm l}=t^{\rm l}_{\rm ff}
=\delta^{-1} t_{\rm turb}$ (Sect.~\ref{sec:vff}). Combining Eq.~(\ref{eq:rhostar})
and Eq.~(\ref{eq:starform}) we get
\begin{equation}
\dot{\Sigma}_{*}=\Phi_{\rm V} \frac{\rho}{t_{\rm turb}}\delta l_{\rm driv}=
\Phi_{\rm V} \delta \rho v_{\rm turb}\ .
\end{equation}
A simple estimate gives:
\begin{equation}
t_{\rm turb} \sim \big( \frac{l_{\rm driv}}{100\ {\rm pc}}\big)
\big( \frac{v_{\rm turb}}{10\ {\rm km\,s}^{-1}} \big) ^{-1} \sim 10\ {\rm Myr}\ {\rm and}
\end{equation}
\begin{equation}
t^{\rm l}_{\rm ff} \sim \sqrt{3\pi/\big(32 G \big( \frac{\rho_{\rm cl}}{500\ {\rm cm}^{-3}}
\big)\big)} \sim 2\ {\rm Myr}\ .
\end{equation}
Since $t_{\rm turb}= \delta t^{\rm l}_{\rm ff}$, this leads to $\delta$=5, which
is precisely the value we find in order to fit the gas properties of the Galaxy
(Sect.~\ref{sec:galaxy}).

\subsection{Energy flux conservation \label{sec:efc}}

Here we assume that the energy input into the ISM due to SNe is transported
practically without loss from the driving length scale to
the dissipative length scale where it is radiated away. Mass is stored into
self-gravitating clouds and eventually used for star formation.
The energy per unit time which is transferred by turbulence is
\begin{equation}
\dot{E} \simeq - \rho \nu \int v_{\rm turb}^{2}/l_{\rm driv}^{2}\,{\rm d}V\ ,
\end{equation}
where the integration is taken over the volume $\int {\rm d}V=V=A\,H$
(Landau \& Lifschitz, 1959).
Thus, the energy flux per unit time and unit area is
\begin{equation}
\frac{\Delta E}{\Delta t\,\Delta A}=- \Sigma \nu
\frac{v_{\rm turb}^{2}}{l_{\rm driv}^{2}}\ .
\label{eq:efcturb}
\end{equation}
This energy loss is balanced by the SNe energy input
\begin{equation}
\dot{E}_{\rm SN}+\dot{E}=0\ .
\label{eq:efc}
\end{equation}
We connect this energy input directly to the star formation rate
(Eq.~(\ref{eq:starform})).
With the assumption of a constant initial mass function (IMF)
independent of the environment one can write
\begin{equation}
\frac{\dot{E}_{\rm SN}}{\Delta A}=\xi\,\dot{\Sigma}_{*}\ .
\end{equation}

The factor of proportionality $\xi$ relates the local SN energy input
to the local star formation rate and is taken to be independent of the radius.
It can be normalized with Galactic
observations by integrating over the Galactic disk:
The number of SN per time is proportional to the Galactic
integrated star formation rate:
\begin{equation}
\dot{N}_{\rm SN}=\tilde{\xi}\,\dot{M}_{*}\ .
\end{equation}
The energy input per surface area and time in Eq.~(\ref{eq:efcturb}) is
\begin{equation}
\frac{\Delta E}{\Delta A \Delta t}=\frac{E^{\rm kin}_{\rm SN} \dot{N}_{\rm SN}}{
\Delta A}=\frac{E^{\rm kin}_{\rm SN} \tilde{\xi} \dot{M}_{*}}{\Delta A}=
\tilde{\xi} E^{\rm kin}_{\rm SN} \dot{\Sigma}_{*}=\xi\,\dot{\Sigma}_{*}\ ,
\end{equation}
where $E^{\rm kin}_{\rm SN}$ is the kinetic energy input from a single SN.
Thus the energy released into the ISM per mass turned into stars is
\begin{equation}
\xi= \frac{\dot{N}_{\rm SN}}{\dot{M}_{*}} E^{\rm kin}_{\rm SN}\ .
\end{equation}

Thornton et al. (1998) have shown by modeling SN explosions in different
environments that the kinetic energy of the remnants is $\sim$10\% of
the total SN energy $E^{\rm tot}_{\rm SN}$ irrespective of the density
and metalicity of the ambient medium. The SN energy input into the ISM
is thus $E^{\rm kin}_{\rm SN} \simeq 10^{50}$~ergs.
The integrated number of SNe type II in the Galaxy is taken to be
$\dot{N}_{\rm SN} \sim 1/40$~yr$^{-1}$ in accordance with
Tammann et al. (1994).
The Galactic star formation rate is taken to be $\dot{M}_{*}$=3~M$_{\odot}$yr$^{-1}$
(Prantzos \& Aubert 1995).
This leads to $\xi \simeq 4.6\,10^{-8}$~(pc/yr)$^{2}$.

\subsection{The molecular fraction}

In order to derive an expression for the molecular fraction of gas in the disk,
we compare the crossing time of the turbulent layer $t_{\rm turb}$ and the
H--H$_{2}$ transition time scale $t_{\rm H_{2}}=\alpha \Phi_{\rm V}/\rho$
(Hollenbach \& Tielens 1997).
We define the molecular fraction here as $f_{\rm mol}=t_{\rm turb}/t_{\rm H_{2}}$.
This allows us to calculate the molecular surface density and total mass that
can be compared with observations.

\section{Results \label{sec:results}}

With $\xi$ fixed and with a given rotation curve our model has only three free
parameters: the constant of the molecule formation rate $\alpha$, the Toomre parameter
$Q$ and the mass accretion rate within the disk $\dot{M}$.
All disk properties can be expressed as functions of $\alpha$, $Q$, $\delta$, $\dot{M}$, $\xi$
$\Omega$, $\partial \Omega / \partial R$, and $R$.
We have solved the set of equations Eq.~(\ref{eq:visc}), Eq.~(\ref{eq:ame}),
Eq.~(\ref{eq:srh}), Eq.~(\ref{eq:pp1})/(\ref{eq:pp2}), Eq.~(\ref{eq:tq}), and Eq.~(\ref{eq:efc}),
for the two cases:
\begin{enumerate}
\item
dominating stellar disk (DSD) ($Q > 1$, S0 and gas deficient galaxies),
\item
self-gravitating gas disk in $z$ direction (SGZ) ($Q \sim 1$, spiral galaxies).
\end{enumerate}
We will use $\Omega'=\partial \Omega / \partial R$ with $\Omega' < 0$
in the region of interest.

\subsection{Dominating stellar disk (DSD) \label{sec:dsd}}

In this case the vertical pressure equilibrium is given by
$p_{\rm grav}=\Sigma \partial \Phi / \partial z$. We make the
approximation $\partial \Phi / \partial z \sim \chi v_{\rm rot}^{2} H^{-1}$
leading to $p_{\rm grav}=\chi \pi G \Sigma_{*} \Sigma$ (see e.g. Binney \& Tremaine 1987).
The factor $\chi$ allows a smooth transition between the SGZ model (Sect.~\ref{sec:sgz})
and the DSD model.

As in Sect.~\ref{sec:sgz} we first give the expression for $H$, $l_{\rm driv}$, $\rho$,
and $\Sigma$ as functions of $v_{\rm turb}$.
\begin{equation}
H = \chi^{-1} R \big(\frac{v_{\rm turb}}{v_{\rm rot}}\big)^{2}
\label{eq:hdsd}
\end{equation}
\begin{equation}
l_{\rm driv}=\sqrt{\frac{3}{2}}\frac{\pi}{8}\,\chi^{\frac{1}{2}}\,G\,\dot{M}\,Q^{\frac{3}{2}}\,
v_{\rm turb}^{-\frac{5}{2}}\,\Omega^{\frac{1}{2}}\,R^{-\frac{1}{2}}\,(-\Omega')^{-1}\ ,
\label{eq:ldrivdsd}
\end{equation}
\begin{equation}
\rho=\pi^{-1}\,\chi\,G^{-1}\,\dot{M}^{-1}\,\Omega^{2}\,\frac{v_{\rm rot}}{v_{\rm turb}}\ ,
\end{equation}
\begin{equation}
\Sigma=\pi^{-1}\,G^{-1}\,Q^{-1}\,v_{\rm turb}\,\Omega\ .
\label{eq:sigmadsd}
\end{equation}

The expression for the turbulent velocity dispersion is
\begin{equation}
v_{\rm turb}=0.87\,\chi^{\frac{4}{15}}G^{\frac{1}{5}}\dot{M}^{\frac{1}{5}}
Q^{\frac{4}{15}}\delta^{-\frac{1}{15}}\xi^{\frac{1}{15}}
R^{\frac{1}{15}}\Omega^{\frac{7}{15}}(-\Omega')^{-\frac{1}{5}}\ .
\label{eq:vturbdsd}
\end{equation}
Inserting Eq.~(\ref{eq:vturbdsd}) into Eq.~(\ref{eq:hdsd}), (\ref{eq:ldrivdsd}),
and \ref{eq:sigmadsd} leads to
\begin{equation}
H=0.75\,\chi^{-\frac{7}{15}}G^{\frac{2}{5}}\dot{M}^{\frac{2}{5}}Q^{\frac{8}{15}}
\delta^{-\frac{2}{15}}
\xi^{\frac{2}{15}}R^{-\frac{13}{15}}\Omega^{-\frac{16}{15}}(-\Omega')^{-\frac{2}{5}}\ ,
\end{equation}
\begin{equation}
l_{\rm driv}=0.69\,\chi^{-\frac{1}{6}}G^{\frac{1}{2}}\dot{M}^{\frac{1}{2}}Q^{\frac{5}{6}}
\delta^{\frac{1}{6}}
\xi^{-\frac{1}{6}}R^{-\frac{2}{3}}\Omega^{-\frac{2}{3}}(-\Omega')^{-\frac{1}{2}}\ ,
\end{equation}
\begin{equation}
\Sigma=0.28\,\chi^{\frac{4}{15}}G^{-\frac{4}{5}}\dot{M}^{\frac{1}{5}}Q^{-\frac{11}{15}}
\delta^{-\frac{1}{15}}
\xi^{\frac{1}{15}}R^{\frac{1}{15}}\Omega^{\frac{22}{15}}(-\Omega')^{-\frac{1}{5}}\ .
\end{equation}
The viscosity $\nu$, the volume filling factor $\phi_{\rm V}$, the star
formation rate $\dot{\Sigma}_{*}$, and the molecular fraction $f_{\rm mol}$
then write
\begin{equation}
\nu=0.58\,\chi^{-\frac{4}{15}}G^{\frac{4}{5}}\dot{M}^{\frac{4}{5}}Q^{\frac{11}{15}}
\delta^{\frac{1}{15}}
\xi^{-\frac{1}{15}}R^{-\frac{16}{15}}\Omega^{-\frac{7}{15}}(-\Omega')^{-\frac{4}{5}}\ ,
\end{equation}
\begin{equation}
\phi_{\rm V}=0.79\,\chi^{-\frac{2}{15}}G^{\frac{2}{5}}\dot{M}^{\frac{2}{5}}Q^{-\frac{2}{15}}
\delta^{-\frac{22}{15}}
\xi^{-\frac{8}{15}}R^{-\frac{8}{15}}\Omega^{\frac{4}{15}}(-\Omega')^{-\frac{2}{5}}\ ,
\end{equation}
\begin{equation}
\dot{\Sigma}_{*}=0.25\,\chi^{\frac{13}{15}}G^{-\frac{3}{5}}\dot{M}^{\frac{2}{5}}
Q^{-\frac{17}{15}}\delta^{-\frac{7}{15}}
\xi^{-\frac{8}{15}}R^{\frac{7}{15}}\Omega^{\frac{49}{15}}(-\Omega')^{-\frac{2}{5}}\ ,
\end{equation}
\begin{equation}
f_{\rm mol}=0.37\,\chi^{\frac{13}{30}}G^{-\frac{13}{10}}\dot{M}^{-\frac{3}{10}}
Q^{-\frac{17}{30}}
\xi^{\frac{7}{30}}R^{\frac{11}{15}}\Omega^{\frac{17}{15}}(-\Omega')^{\frac{3}{10}}
\alpha^{-1}\ .
\end{equation}
Setting $\Omega' \simeq -\Omega/R$ and $v_{\rm rot}=\Omega \, R$ leads to the following
simple radial dependences:
$v_{\rm turb} = (\chi\,v_{\rm rot})^{\frac{4}{15}}$,
$H \propto (\chi\,v_{\rm rot})^{-\frac{7}{15}}\Omega^{-1}$,
$l_{\rm driv} \propto (\chi\,v_{\rm rot})^{-\frac{1}{6}}\Omega^{-1}$,
$\Sigma \propto (\chi\,v_{\rm rot})^{\frac{4}{15}}\Omega$,
$\nu \propto (\chi\,v_{\rm rot})^{-\frac{4}{15}}\Omega^{-1}$,
$\phi_{\rm V} = (\chi\,v_{\rm rot})^{-\frac{2}{15}}$,
$\dot{\Sigma}_{*} \propto (\chi\,v_{\rm rot})^{\frac{13}{15}}\Omega^{2}$,
and $f_{\rm mol} \propto (\chi\,v_{\rm rot})^{\frac{13}{30}}\Omega$.

\begin{figure}
        \resizebox{\hsize}{!}{\includegraphics{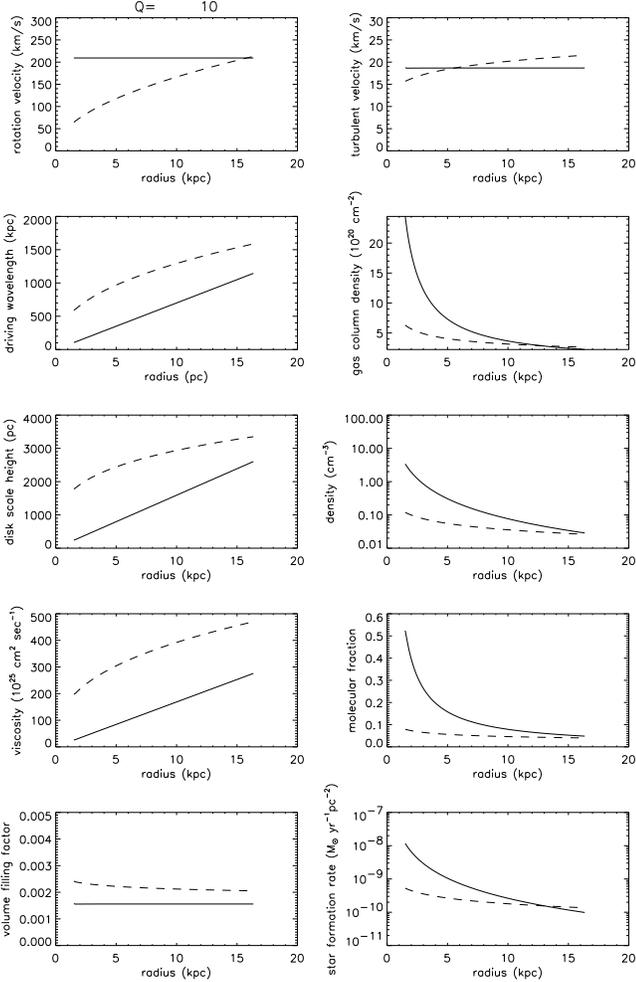}}
        \caption{ \label{fig:graphs_dsd_sn_l0_new}
        Disk parameters for the case of the DSD model.
        Solid lines: constant rotation curve. Dashed line: rising
        rotation curve $v_{\rm rot} \propto \sqrt{R}$.
        }
\end{figure}

We found that for $Q$=1 and $\chi \sim 0.05$ there is a smooth transition between
the SGZ and DSD model, thus
$\partial \Phi / \partial z \simeq 0.05 v_{\rm rot}^{2} R^{-1}$.
Fig.~\ref{fig:graphs_dsd_sn_l0_new} shows the radial dependences of the above
parameters for (i) a constant rotation curve and (ii) for a rising
rotation curve. We have included an arbitrary rising rotation curve to show
the general effects of $(-\Omega')$ on the disk properties.
We chose $\delta$ in a way to fit the observed volume filling factor of
molecular clouds (Table~\ref{tab:tab1}).
For this model $\chi=0.05$, $Q=10$, $\delta=5$, $\dot{M}=10^{-1}$~M$_{\odot}$yr$^{-1}$, and
$\alpha =3\,10^{7}$~yr\,M$_{\odot}$pc$^{-3}$.
The rising rotation curve leads to a bigger driving
length scale, disk height, viscosity, and volume filling factor.
On the other hand, it leads to a smaller surface density, density, molecular
fraction, and star formation rate in the inner disk than for a constant rotation curve.
The total gas mass is $M_{\rm gas} \sim 1.3/1.0\,10^{9}$~M$_{\odot}$, the total
molecular mass is $M_{\rm gas} \sim 1.7/0.5\,10^{8}$~M$_{\odot}$,
and the total star formation is $\dot{M}_{*} \sim 0.2/0.1$~M$_{\odot}$yr$^{-1}$
for the model with a flat and a rising rotation curve, respectively.
This model applies to gas poor spiral and S0 galaxies.

For this model we have used $\dot{\Sigma}_{*}=\dot{\rho}_{*}\,l_{\rm driv}$.
This is only valid if $H/l_{\rm driv} > 1$ with
\begin{equation}
\frac{H}{l_{\rm driv}}=1.09\,\chi^{-\frac{3}{10}}G^{-\frac{1}{10}}\dot{M}^{-\frac{1}{10}}
Q^{-\frac{3}{10}}\delta^{-\frac{3}{10}}\xi^{\frac{3}{10}}v_{\rm rot}^{-\frac{3}{10}}\ .
\end{equation}
For $v_{\rm rot}$=220~km\,s$^{-1}$, $\dot{M}=10^{-2}$~M$_{\odot}$yr$^{-1}$,
and $\chi=0.05$ this translates into the requirement $Q<1440$.

In the case of $H/l_{\rm driv} < 1$, we set $\dot{\Sigma}_{*}=\dot{\rho}_{*}\,H$.
Whereas Eq.~(\ref{eq:hdsd}) -- Eq.~(\ref{eq:sigmadsd}) do not change, the
turbulent velocity becomes
\begin{equation}
v_{\rm turb}=0.87\,\chi^{\frac{5}{21}}G^{\frac{4}{21}}\dot{M}^{\frac{4}{21}}
Q^{\frac{5}{21}}\delta^{-\frac{2}{21}}\xi^{\frac{2}{21}}
R^{\frac{1}{21}}\Omega^{\frac{9}{21}}(-\Omega')^{-\frac{4}{21}}\ .
\end{equation}
The exponents are only slightly different from those of Eq.~(\ref{eq:vturbdsd}).

\subsection{Self-gravitating gas disk in $z$ direction (SGZ) \label{sec:sgz}}

We first give the expressions for the disk height $H$, the turbulent driving
length scale $l_{\rm driv}$, the mean density in the disk plane $\rho$, and
the disk surface density $\Sigma$ as functions of the Toomre parameter $Q$,
the mass accretion rate $\dot{M}$, the angular velocity $\Omega$, the disk radius
$R$, the radial derivative of the angular velocity $\Omega'$, and
the turbulent velocity dispersion $v_{\rm turb}$.
\begin{equation}
H=\frac{Q\,v_{\rm turb}}{\Omega}
\label{eq:hsgz}
\end{equation}
\begin{equation}
l_{\rm driv}=\sqrt{\frac{3}{2}}\frac{\pi\,G\,\dot{M}\,Q}{8\,R\,v_{\rm turb}^{2}
(-\Omega')}
\label{eq:ldrivsgz}
\end{equation}
\begin{equation}
\rho=\frac{\Omega^{2}}{\pi\,G\,Q^{2}}
\label{eq:rhosgz}
\end{equation}
\begin{equation}
\Sigma=\frac{v_{\rm turb}\,\Omega}{\pi\,G\,Q}
\label{eq:sigmasgz}
\end{equation}
The expression for the turbulent velocity dispersion is
\begin{equation}
v_{\rm turb}=0.82\,G^{\frac{3}{11}}\dot{M}^{\frac{3}{11}}\delta^{-\frac{1}{11}}
\xi^{\frac{1}{11}}R^{-\frac{3}{11}}
\Omega^{\frac{3}{11}}(-\Omega')^{-\frac{3}{11}}\ .
\label{eq:vturbsgz}
\end{equation}

Since $Q=(v_{\rm turb}/v_{\rm rot})(M_{\rm tot}/M_{\rm gas})$ (Eq.~(\ref{eq:tqint}))
and the turbulent velocity dispersion $v_{\rm turb}$ is independent of $Q$,
the Toomre parameter is only a measure of the ratio between the total
enclosed and the total gas mass of the galaxy.

Using Eq.~(\ref{eq:nuu}) together with the pressure equilibrium in $z$ direction
the viscosity intermittence factor $\gamma$ of Eq.~(\ref{eq:visc}) can be calculated
\begin{equation}
\gamma = \frac{H}{l_{\rm driv}} \sqrt{\phi_{\rm V}}=\sqrt{\frac{32}{3\pi^{2}}} \simeq 1\ .
\end{equation}

Thus, in the SGZ model the viscosity reads
\begin{equation}
\nu = \sqrt{\phi_{\rm V}}\,v_{\rm turb}\,H = v_{\rm turb}\,l_{\rm driv}\ .
\end{equation}

Inserting Eq.~(\ref{eq:vturbsgz}) into Eq.~(\ref{eq:hsgz}), (\ref{eq:ldrivsgz}), and
(\ref{eq:sigmasgz}) leads to
\begin{equation}
H=0.82\,G^{\frac{3}{11}}\dot{M}^{\frac{3}{11}}Q\delta^{-\frac{1}{11}}
\xi^{\frac{1}{11}}R^{-\frac{3}{11}}\Omega^{-\frac{8}{11}}(-\Omega')^{-\frac{3}{11}}\ ,
\end{equation}
\begin{equation}
l_{\rm driv}=0.71\,G^{\frac{5}{11}}\dot{M}^{\frac{5}{11}}Q\delta^{\frac{2}{11}}
\xi^{-\frac{2}{11}}R^{-\frac{5}{11}}\Omega^{-\frac{6}{11}}(-\Omega')^{-\frac{5}{11}}\ ,
\end{equation}
\begin{equation}
\Sigma=0.26\,G^{-\frac{8}{11}}\dot{M}^{\frac{3}{11}}Q^{-1}\delta^{-\frac{1}{11}}
\xi^{\frac{1}{11}}R^{-\frac{3}{11}}\Omega^{\frac{14}{11}}(-\Omega')^{-\frac{3}{11}}\ .
\label{eq:sigmasgz1}
\end{equation}
For the viscosity $\nu$, the volume filling factor $\phi_{\rm V}$, the star
formation rate $\dot{\Sigma}_{*}$, and the molecular fraction $f_{\rm mol}$
one obtains
\begin{equation}
\nu=0.61\,G^{\frac{8}{11}}\dot{M}^{\frac{8}{11}}Q\delta^{\frac{1}{11}}\xi^{-\frac{1}{11}}
R^{-\frac{8}{11}}\Omega^{-\frac{3}{11}}(-\Omega')^{-\frac{8}{11}}\ ,
\end{equation}
\begin{equation}
\phi_{\rm V}=0.81\,G^{\frac{4}{11}}\dot{M}^{\frac{4}{11}}\delta^{-\frac{16}{11}}
\xi^{-\frac{6}{11}}R^{-\frac{4}{11}}\Omega^{\frac{4}{11}}(-\Omega')^{-\frac{4}{11}}\ ,
\end{equation}
\begin{equation}
\dot{\Sigma}_{*}=0.21\,G^{-\frac{4}{11}}\dot{M}^{\frac{7}{11}}Q^{-2}\delta^{-\frac{6}{11}}
\xi^{-\frac{5}{11}}R^{-\frac{7}{11}}\Omega^{\frac{29}{11}}(-\Omega')^{-\frac{7}{11}}\ ,
\label{eq:starsgz1}
\end{equation}
\begin{equation}
f_{\rm mol}=0.34\,G^{-\frac{13}{11}}\dot{M}^{-\frac{2}{11}}Q^{-1}\xi^{\frac{3}{11}}
R^{\frac{2}{11}}\Omega^{\frac{9}{11}}(-\Omega')^{\frac{2}{11}}\alpha^{-1}\ .
\end{equation}

Setting $\Omega' \simeq -\Omega/R$ leads to the following simple radial scalings:
$v_{\rm turb} = const.$, $H \propto \Omega^{-1}$, $l_{\rm driv} \propto \Omega^{-1}$,
$\Sigma \propto \Omega$, $\nu \propto \Omega^{-1}$, $\phi_{\rm V} = const.$,
$\dot{\Sigma}_{*} \propto \Omega^{2}$, and $f_{\rm mol} \propto \Omega$.

\begin{figure}
        \resizebox{\hsize}{!}{\includegraphics{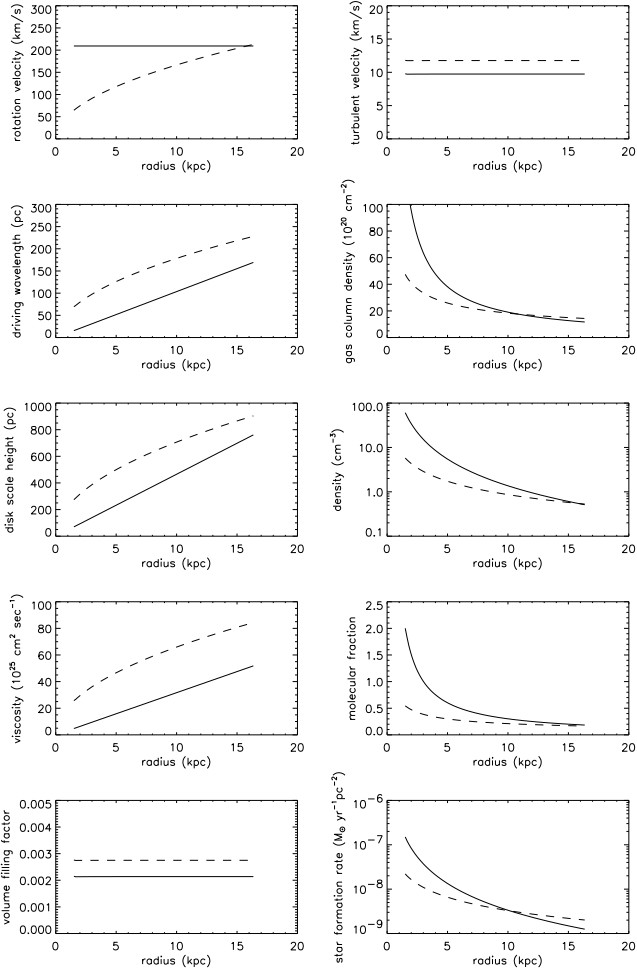}}
        \caption{ \label{fig:graphs_sgz_sn}
        Disk parameters for the SGZ model.
        Solid lines: disk with constant rotation curve. Dashed line: disk with rising
        rotation curve $v_{\rm rot} \propto \sqrt{R}$.
        }
\end{figure}

Fig.~\ref{fig:graphs_sgz_sn} shows the radial dependences of these
parameters for (i) a constant rotation curve and (ii) for a rising
rotation curve. We have again included a rising rotation curve to show
the general effects of $(-\Omega')$ on the disk properties.
For this model $Q=1$, $\delta=5$, $\dot{M}=10^{-1}$~M$_{\odot}$yr$^{-1}$, and
$\alpha = 3\,10^{7}$~yr\,M$_{\odot}$pc$^{-3}$ in order to fit Galactic observations.
The rising rotation curve leads to a bigger driving
length scale, disk height, viscosity, and volume filling factor.
On the other hand, it leads to a smaller surface density, density, molecular
fraction, and star formation rate mainly in the inner disk.
The fraction between the gas density, and star formation rate
for a constant and those for a rising rotation curve is about 5--10 in the central
part of the galaxy. This fraction is about 3--5 for the surface density, driving length scale,
disk height, and viscosity. The difference in the velocity dispersion
and the volume filling factor is 20\% and 30\% respectively.
The total gas mass for both models is $M_{\rm gas} \sim 6.8\,10^{9}$~M$_{\odot}$,
the total molecular mass is $M_{\rm mol} \sim 3.0\,10^{9}$~M$_{\odot}$,
and the total star formation is $\dot{M}_{*} \sim 2.6$~M$_{\odot}$yr$^{-1}$
in good agreement with Galactic observations (for the gas see, e.g.
Kulkarni \& Heiles 1988; for the star formation rate see, e.g.
Prantzos \& Aubert 1995).

\section{Comparison with the fully gravitational model of Vollmer \& Beckert (2002) \label{sec:fullgrav}}

In Vollmer \& Beckert (2002) we have analytically investigated the equilibrium
state of a turbulent clumpy gas disk that consists of distinct
self-gravitating clouds, which are embedded
in a low density medium, and evolve in the fixed gravitational potential
of the galaxy.  Gravitational cloud--cloud interactions in the
disk give rise to an effective viscosity and allows the transport of
angular momentum and mass in the gas disk. In this scenario turbulence
is assumed to be generated by instabilities involving self-gravitation
and to be maintained by the energy input, which is provided by differential
rotation of the disk and mass transfer to smaller galactic radii via
cloud--cloud interactions.
Both the energy source and dissipation process for turbulence in these
disks are due to gravity. Therefore, we call this model the fully
gravitational model.

\subsection{The viscosity prescription}

In the fully gravitational model the viscosity prescription is
\begin{equation}
\nu=Re^{-1}\,v_{\rm turb}\,H\ .
\label{eq:revisc}
\end{equation}
In Vollmer \& Beckert (2002) we have used a slightly different expression
for the energy flux transferred by turbulence to smaller length scales:
$\dot{E}=\rho \nu v_{\rm turb}^{2} / l_{\rm driv} = \Sigma \nu v_{\rm turb}^{2}
/ (H\,l_{\rm driv})$ instead of $\dot{E}=\Sigma \nu v_{\rm turb}^{2} /
l_{\rm driv}^{2}$. Thus we had already assumed $l_{\rm driv}=H$.
This lead to exponents of $Q$ that were twice that of the model using
Eq.~(\ref{eq:efcturb}).
Therefore, the volume filling factor in the case of a turbulent non-Kolmogorov energy
spectrum ($D=2$)
of the form $E(k) \propto k^{-2}$ in the framework of Vollmer \& Beckert (2002)
using Eq.~(\ref{eq:efcturb}) is
\begin{equation}
\phi_{\rm V} \simeq Q^{-2}\,Re^{-2}\ .
\end{equation}

For $Q=1$ one gets $\phi_{\rm V} = Re^{-2}$.
In the present model including SNe, Eq.~(\ref{eq:visc}) reads for the viscosity
\begin{equation}
\nu=\sqrt{\phi_{\rm V}}\,v_{\rm turb}\,\delta\,H\ .
\end{equation}
Thus, for $Q=1$ and $\delta=1$ the viscosity prescriptions of
both models are equivalent.

\subsection{The star formation rate}

In Vollmer \& Beckert (2002) we suggested a star formation law of the form
\begin{equation}
\dot{\Sigma}_{*}=Re^{-1}\,\Sigma\,\Omega\ .
\end{equation}
This has to be compared with the star formation rate in the present model,
which reads
(see Sect.~\ref{sec:starform}).
\begin{equation}
\dot{\Sigma}_{*}=\sqrt{\phi_{\rm V}}\,\rho\,l_{\rm driv}\,(t^{\rm H}_{\rm ff})^{-1}\ .
\end{equation}
In the case of the SGZ model
$\rho=(\pi\,G\,Q^{2})^{-1} \Omega^{2}$ (Eq.~(\ref{eq:rhosgz})).
Thus, for $Q=1$, $(t^{\rm H}_{\rm ff})^{-1} \simeq \Omega$ leading to a star formation
rate of the form
\begin{equation}
\dot{\Sigma}_{*}=\sqrt{\phi_{\rm V}}\,\rho\,l_{\rm driv}\,\Omega\ .
\end{equation}
Since in the fully gravitational model $l_{\rm driv}=H$, the prescription
for the star formation rate of both models are equivalent for
$Q=1$ disks.

\subsection{Energy flux conservation}

The main difference between the fully gravitational model and the
model including SNe lies in the energy flux
conservation equation. The energy transferred by turbulence to smaller
length scales is balanced by the energy input through SNe in our present
model, whereas it is balanced by the differential rotation and mass inflow
in the fully gravitational model (Vollmer \& Beckert 2002).

\subsection{Results}

If one approximates $\Omega' \sim - \Omega / R$, the radial scalings of
all disk properties is the same for both SGZ models,
i.e. $v_{\rm turb} = const.$, $H \propto \Omega^{-1}$,
$l_{\rm driv} \propto \Omega^{-1}$,
$\Sigma \propto \Omega$, $\nu \propto \Omega^{-1}$, $\phi_{\rm V} = const.$,
$\dot{\Sigma}_{*} \propto \Omega^{2}$, and $f_{\rm mol} \propto \Omega$.

We therefore conclude that the fully gravitational model (Vollmer \& Beckert 2002)
and the here presented model including SNe lead to the same radial
dependences of the disk properties in the case of $Q=1$
with $Re=\phi_{\rm V}^{-1}$. A quantitative comparison will be
made in Sect.~\ref{sec:galaxy}.

For the comparison with previous models of turbulent, self-gravitating gas
disks we refer to Sect.~5.1 in Vollmer \& Beckert (2002).

\section{SN remnant size as driving wavelength (SNR) \label{sec:snr}}

In Sect.~\ref{sec:efc} the energy flux conservation equation determines the
turbulent driving wavelength $l_{\rm driv}$. Alternatively, one can assume
that only the interaction of SN bubbles leads to an effective turbulent viscosity
that transports angular momentum. In this case the driving wavelength is twice the radius
of the SN remnant at a time when the shock thermalizes.
In the following we will call this model the {\sc SNR} model.
Following Dorfi (1993) the final radius of a SN remnant is
given by
\begin{equation}
R_{\rm SNR}=l_{\rm SN} E_{51}^{\frac{11}{35}} n_{0}^{-\frac{13}{35}}\ {\rm pc}\ ,
\label{eq:rsnr}
\end{equation}
where $l_{\rm SN}$ is the characteristic radius for $n_{0}$=1~cm$^{-3}$,
$E_{51}=E/(10^{51}$~erg), and $n_{0}$ is the average local density in cm$^{-3}$.
We set $l_{\rm SN}$=64~pc (Dorfi 1993) and replace Eq.~(\ref{eq:efc}) by
$l_{\rm driv}=2\,R_{\rm SNR}$.
This results in
\begin{equation}
v_{\rm turb}=0.56\,G^{\frac{11}{35}}Q^{\frac{9}{70}}E_{51}^{-\frac{11}{70}}l_{\rm SN}^{-\frac{1}{2}} \dot{M}^{\frac{1}{2}} R^{-\frac{1}{2}} \Omega^{\frac{13}{35}} (-\Omega')^{-\frac{1}{2}}\ ,
\end{equation}
\begin{equation}
H=0.56\,G^{\frac{11}{35}}Q^{\frac{79}{70}}E_{51}^{-\frac{11}{70}}l_{\rm SN}^{-\frac{1}{2}} \dot{M}^{\frac{1}{2}} R^{-\frac{1}{2}} \Omega^{-\frac{22}{35}} (-\Omega')^{-\frac{1}{2}}\ ,
\end{equation}
\begin{equation}
l_{\rm driv}=1.53\,G^{\frac{13}{35}}Q^{\frac{26}{35}} E_{51}^{\frac{11}{35}}l_{\rm SN} \Omega^{-\frac{26}{35}}\ ,
\end{equation}
\begin{equation}
\rho=\frac{\Omega^{2}}{\pi G Q^{2}}\ ,
\label{eq:romeg}
\end{equation}
\begin{equation}
\Sigma=0.18\,G^{-\frac{24}{35}}Q^{-\frac{61}{70}}E_{51}^{-\frac{11}{70}}l_{\rm SN}^{-\frac{1}{2}} \dot{M}^{\frac{1}{2}} R^{-\frac{1}{2}} \Omega^{\frac{48}{35}} (-\Omega')^{-\frac{1}{2}}\ ,
\end{equation}
\begin{equation}
\nu=0.89\,G^{\frac{24}{35}}Q^{\frac{61}{70}}E_{51}^{\frac{11}{70}}l_{\rm SN}^{\frac{1}{2}} \dot{M}^{\frac{1}{2}} R^{-\frac{1}{2}} \Omega^{-\frac{13}{35}} (-\Omega')^{-\frac{1}{2}}\ ,
\end{equation}
\begin{equation}
\Phi_{\rm V}=8.05\,G^{\frac{4}{35}}Q^{-\frac{27}{35}}\delta^{-2}E_{51}^{\frac{33}{35}}l_{\rm SN}^{3} \dot{M}^{-1} R \Omega^{-\frac{8}{35}} (-\Omega')\ ,
\end{equation}
\begin{equation}
\dot{\Sigma}_{*}=1.44\,G^{-\frac{4}{7}}Q^{-\frac{37}{14}}\delta^{-1}E_{51}^{\frac{11}{14}}l_{\rm SN}^{\frac{5}{2}} \dot{M}^{-\frac{1}{2}} R^{\frac{1}{2}} \Omega^{\frac{15}{7}} (-\Omega')^{\frac{1}{2}}\ ,
\end{equation}
\begin{equation}
f_{\rm mol}=0.11\,G^{-\frac{37}{35}}Q^{-\frac{43}{70}}\delta E_{51}^{-\frac{33}{70}} l_{\rm SN}^{-\frac{3}{2}} \dot{M}^{\frac{1}{2}} R^{-\frac{1}{2}} \Omega^{\frac{39}{35}} (-\Omega')^{-\frac{1}{2}} \alpha^{-1}\ .
\end{equation}
The intermittence factor $\gamma$ (Eq.~(\ref{eq:intfactor})) is in this case $\gamma \sim 1$
as for the SGZ model.

Setting $\Omega' \simeq -\Omega/R$ leads to the following radial scalings:
$v_{\rm turb} \propto \Omega^{-9/70}$, $H \propto \Omega^{-79/70}$, $l_{\rm driv} \propto
\Omega^{-26/35}$, $\Sigma \propto \Omega^{61/70}$, $\nu \propto \Omega^{-61/70}$,
$\Phi_{\rm V} \propto \Omega^{27/35}$, $\dot{\Sigma}_{*} \propto \Omega^{37/14}$, and
$f_{\rm mol} \propto \Omega^{43/70}$. The exponents are quite close to those of
the SGZ model.

\begin{figure}
        \resizebox{\hsize}{!}{\includegraphics{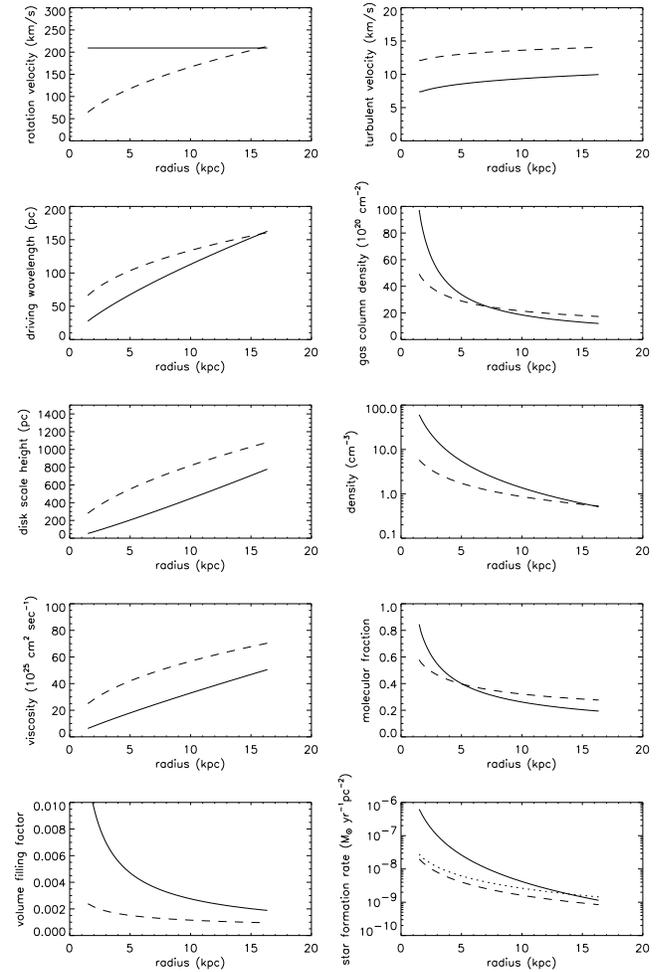}}
        \caption{ \label{fig:graphs_sgz_sn_sedov}
        Disk parameters for the case of SNR model.
        Solid lines: disk with constant rotation curve. Dashed line: disk with rising
        rotation curve $v_{\rm rot} \propto \sqrt{R}$.
        }
\end{figure}

We assume $\dot{M}=10^{-1}$~M$_{\odot}$yr$^{-1}$, $Q=1$, $\delta=5$,
and $\alpha =2\,10^{7}$~yr\,M$_{\odot}$pc$^{-3}$.
Fig.~\ref{fig:graphs_sgz_sn_sedov} shows these parameters for a constant and a
rising rotation curve.
In comparison to the SGZ model, the turbulent
velocity dispersion and the volume filling factor are not constant with radius. For a
constant rotation curve the latter increases by a factor 4 from the outer disk to the center of
the galaxy. All other quantities are similar to those of the SGZ model.

The total gas mass for both rotation curves is $M_{\rm gas} \sim 6.3\,10^{9}$~M$_{\odot}$
and the total molecular gas mass is $M_{\rm mol} \sim 2.0\,10^{8}$~M$_{\odot}$.
The total star formation rate is $\dot{M}_{*} \sim 0.8$~M$_{\odot}$yr$^{-1}$ for the
rising rotation curve and $\dot{M}_{*} \sim 5.7$~M$_{\odot}$yr$^{-1}$ for the
constant rotation curve.

The star formation rate for a constant rotation curve is a factor 2 higher than the
observed Galactic value.
Alternatively, if one drops Eq.~(\ref{eq:starform}) in the initial equations,
one can deduce the star formation rate a posteriori with the help of the energy flux
conservation equation (Eq.~(\ref{eq:efc})):
\begin{equation}
\dot{\Sigma}_{*}=\xi^{-1} \Sigma \nu \big(\frac{v_{\rm turb}}{l_{\rm driv}}\big) ^{2}
\label{eq:sfrefc}
\end{equation}
This is shown for the constant rotation curve as a dotted line in the panel of the star
formation rate in Fig.~\ref{fig:graphs_sgz_sn_sedov}.
Since $\Sigma \nu \sim \dot{M}/(3\,\pi)$ and $v_{\rm turb}=const.$ Eq.~(\ref{eq:sfrefc})
yields $\dot{\Sigma}_{*} \propto l_{\rm driv}^{-2}$. Now, with $l_{\rm driv} \propto
\rho^{-13/35}$ (Eq.~(\ref{eq:rsnr})), $\rho \propto \Omega^{2}$ (Eq.~(\ref{eq:romeg})),
and $\Sigma \propto \Omega^{61/70}$ one obtains $\dot{\Sigma}_{*} \propto \rho^{0.75}
\propto \Omega^{1.5} \propto \Sigma^{1.7}$, which is the observed Schmidt law.
In this case the characteristic timescale for star formation is no longer the
free fall time but it is proportional to the viscous timescale.
The total star formation rate in this case is $\dot{M}_{*} \sim 1.8$~M$_{\odot}$yr$^{-1}$.

A possible remedy would be to use the star formation prescription of Eq.~(\ref{eq:sfrefc})
and assume a smaller $\xi$, i.e. a smaller fraction of the total SN energy that
goes into turbulent motions. With 4\% of the total SN energy in the form of kinetic
energy input as suggested by Dorfi (1993) would raise the total star formation rate
to $\dot{M}_{*} \sim 3.3$~M$_{\odot}$yr$^{-1}$.

\section{Application to the Galaxy \label{sec:galaxy}}

In this Section we will investigate how close the simple analytical models of
Sect.~\ref{sec:sgz} and \ref{sec:snr} can get to the observations of the Galaxy.

\subsection{Self-gravitating disk in $z$ direction (SGZ) \label{sec:galsgz}}

\begin{figure}
        \resizebox{\hsize}{!}{\includegraphics{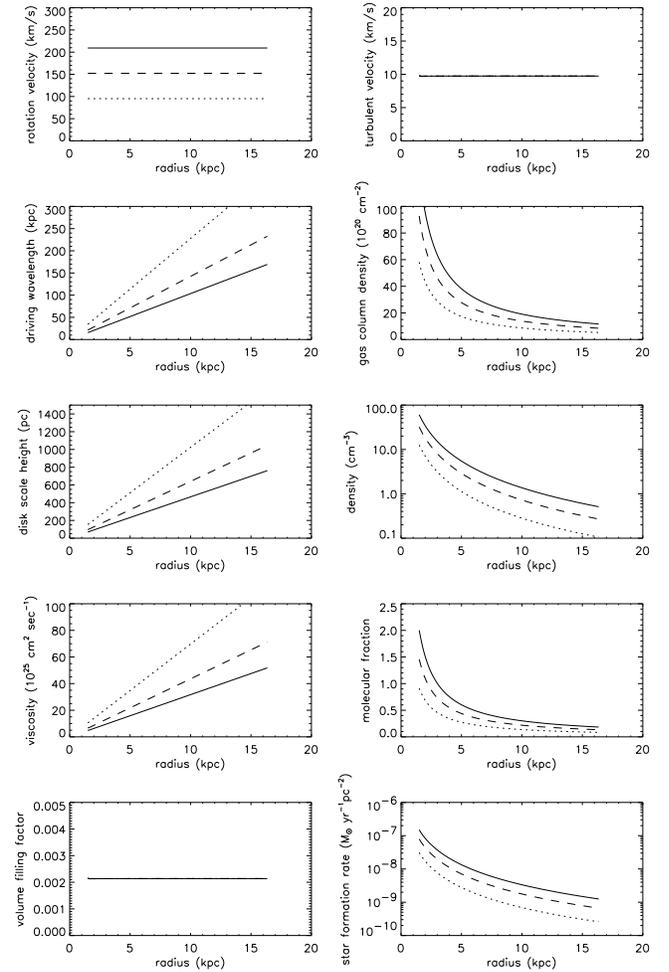}}
        \caption{ \label{fig:graphs_all_sn_new}
        Radial profiles of disk parameters for the case of the SGZ model
        for three different rotation velocities
        (solid lines: $v_{\rm rot}$=250~km\,s$^{-1}$, dashed lines:
        175~km\,s$^{-1}$, dotted lines: 100~km\,s$^{-1}$).
        }
\end{figure}

We use $Q=1.0$, $\delta=5$, $\alpha=3\,10^{7}$~yr\,M$_{\odot}$\,pc$^{-3}$,
$\dot{M}=10^{-1}$~M$_{\odot}$\,yr$^{-1}$, and $\xi=4.6\,10^{-8}$~(pc/yr)$^{2}$.
Fig.~\ref{fig:graphs_all_sn_new} shows the derived turbulent velocity,
driving length scale, gas surface density, disk scale height, gas density,
viscosity, molecular fraction, volume filling factor, and star formation rate
for three different rotation velocities.
For the model of the Galaxy we adopt $v_{\rm rot}=220$~km\,s$^{-1}$.
This results in a total gas mass of $M_{\rm gas} \sim 6.8\,10^{9}$~M$_{\odot}$,
a total molecular mass of $M_{\rm mol} \sim 3.3\,10^{9}$~M$_{\odot}$,
a total atomic gas mass of $M_{\rm HI} \sim 3.5\,10^{9}$~M$_{\odot}$,
and a total star formation rate of $\dot{M}_{*} \sim 2.6$~M$_{\odot}$\,yr$^{-1}$.

The local density $\rho \sim 2$~cm$^{-3}$ and the local surface density
of $\Sigma \sim 2\,10^{21}$~cm$^{-2}$ at the solar radius (Binney \& Tremaine 1987)
are well fitted by our model. At the solar radius the disk scale is
$H \sim 400$~pc, the turbulent driving scale length is $l_{\rm driv} \sim90$~pc,
and the effective viscosity is $\nu \sim 27\,10^{25}$~cm$^{2}$\,s$^{-1}$.
The driving wavelength is comparable to the generally accepted
value of 50~pc$\leq l_{\rm driv} \leq$150~pc (see e.g. Ruzmaikin et al. 1988).
Moreover, the derived star formation rate and gas surface
density are comparable to those observed for the Galaxy.

The constant volume filling factor of $\phi_{\rm V} \sim 2\,10^{-3}$
gives an equivalent Reynolds number of the fully gravitational model
(Vollmer \& Beckert 2002) of $Re = 22$ for a non-Kolmogorov spectrum ($D=2$).
In Vollmer \& Beckert (2002) we
used $Q=1$, $\dot{M}=10^{-2}$~M$_{\odot}$\,yr$^{-1}$,
$\alpha=10^{7}$~yr\,M$_{\odot}$\,pc$^{-3}$ and $Re=50$. All these parameters
are within a factor 2 consistent with the values described above.
This again shows the equivalence of the two models.

\subsection{SN remnant size as driving length scale (SNR)}

\begin{figure}
        \resizebox{\hsize}{!}{\includegraphics{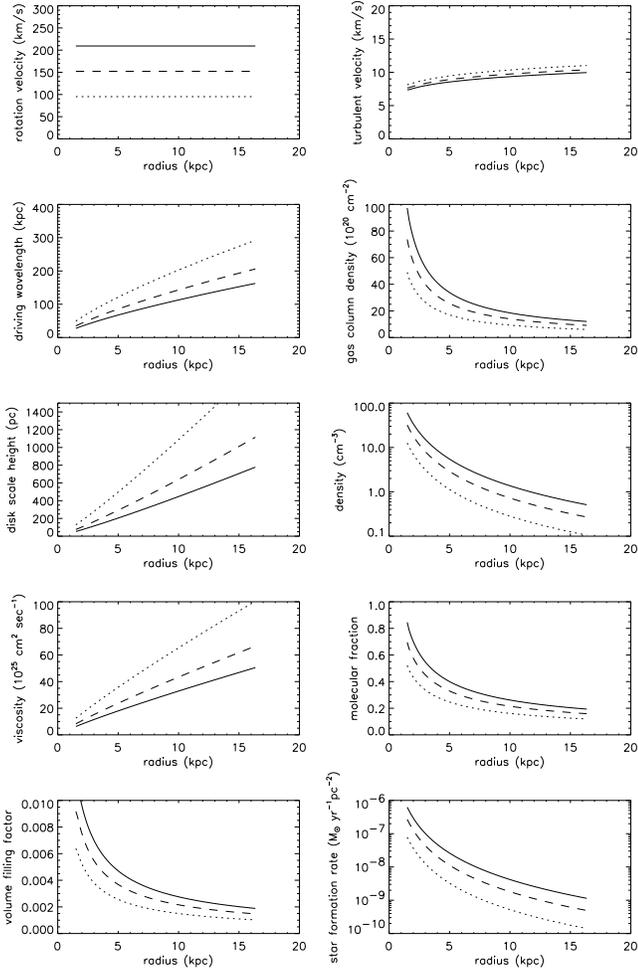}}
        \caption{ \label{fig:graphs_all_sn_new_sedov}
        Radial profiles of disk parameters for the case of
        the SNR model for three different rotation velocities
        (solid lines: $v_{\rm rot}$=250~km\,s$^{-1}$, dashed lines:
        175~km\,s$^{-1}$, dotted lines: 100~km\,s$^{-1}$).
        }
\end{figure}

For this model we use $Q=1$, $\delta=5$, $\alpha=2\,10^{7}$~yr\,M$_{\odot}$\,pc$^{-3}$,
$\dot{M}=10^{-1}$~M$_{\odot}$\,yr$^{-1}$, and
$\xi=4.6\,10^{-8}$~(pc/yr)$^{2}$. The derived disk properties can be seen in
Fig.~\ref{fig:graphs_all_sn_new_sedov}.
For a rotation velocity of $v_{\rm rot}=220$~km\,s$^{-1}$, we obtain the
following results:
total gas mass $M_{\rm gas} \sim 6.3\,10^{9}$~M$_{\odot}$,
total molecular mass of $M_{\rm mol} \sim 2.0\,10^{9}$~M$_{\odot}$,
a total atomic gas mass of $M_{\rm HI} \sim 4.3\,10^{9}$~M$_{\odot}$,
and the total star formation rate given in Sect.~\ref{sec:snr}.

The driving wavelength at the solar radius is $l_{\rm driv}$=100~pc.
The other local quantities ($\Sigma$, $\rho$, $H$) are comparable to those
of Sect.~\ref{sec:galsgz}. As already mentioned the star formation
rate calculated using Eq.~(\ref{eq:starform}) is somewhat too high and that
calculated using the energy flux conservation (Eq.~(\ref{eq:sfrefc})) is somewhat too low
compared to the observed value.

\section{A realistic gravitational potential \label{sec:realpot}}

In this Section we use an analytic gravitational potential $\Phi_{\rm g}$
given by Allen \& Santill\'an (1991). It consists of a disk, bulge, and halo
component. We use the following values for the parameters of Allen \&
Santill\'an (1991):
\begin{itemize}
\item
bulge: $M_{1}=1.41\,10^{10}$~M$_{\odot}$, $b_{1}=387$~pc,
\item
disk: $M_{2}=8.56\,10^{10}$~M$_{\odot}$, $a_{2}=5318$~pc, $b_{2}=250$~pc,
\item
halo: $M_{1}=1.07\,10^{11}$~M$_{\odot}$, $a_{3}=12$~kpc.
\end{itemize}

Vertical pressure equilibrium is described by
\begin{equation}
\rho\,v_{\rm turb}^{2} = \Sigma\,(\pi\,G\,\Sigma + \frac{\partial \Phi_{\rm g}}
{\partial z})\ .
\end{equation}

\subsection{Self-gravitating disk in $z$ direction \label{sec:numsgz}}

We adopt $Q=1.0$, $\delta=5$, $\dot{M}=5\,10^{-2}$~M$_{\odot}$\,yr$^{-1}$,
and $\alpha=4\,10^{7}$~yr\,M$_{\odot}$\,pc$^{-3}$.

\begin{figure}
        \resizebox{\hsize}{!}{\includegraphics{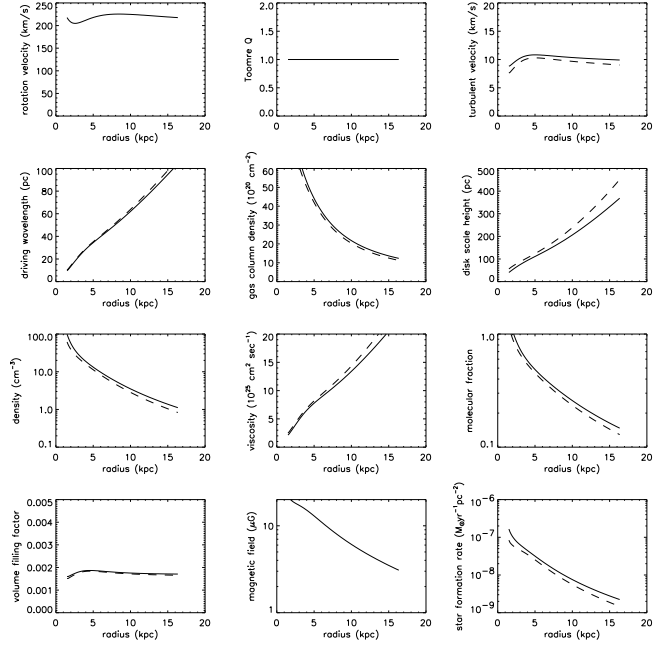}}
        \caption{ \label{fig:numerics_gal_new}
        SGZ model.
        Solid lines: radial profiles of disk parameters for a realistic gravitational
        potential. Dashed lines: disk properties including magnetic and thermal
        pressure.
        }
\end{figure}

The resulting radial profiles for the disk parameters are shown in
Fig.~\ref{fig:numerics_gal_new}.
It turns out that the gas disk is close to the stage of self-gravitation in $z$
direction, but is still dominated by the gravitational potential of
disk, bulge, and halo.
The total gas mass is $M_{\rm tot}=7.5\,10^{9}$~M$_{\odot}$, the total
molecular mass is  $M_{\rm mol}=2.8\,10^{9}$~M$_{\odot}$, and
the star formation rate is $\dot{M}_{*}=5.0$~M$_{\odot}$\,yr$^{-1}$.

For $R > 5$~kpc the gas surface density has approximately a 1/$R$ profile.
The disk is flaring with an increasing flaring angle for increasing $R$.
The volume filling factor is still approximately constant with radius.
The star formation rate can be described by an exponential with
a scale length of $\sim$3.5~kpc for radii between 3 and 10~kpc.

\subsection{SN remnant size as driving length scale (SNR)}

\begin{figure}
        \resizebox{\hsize}{!}{\includegraphics{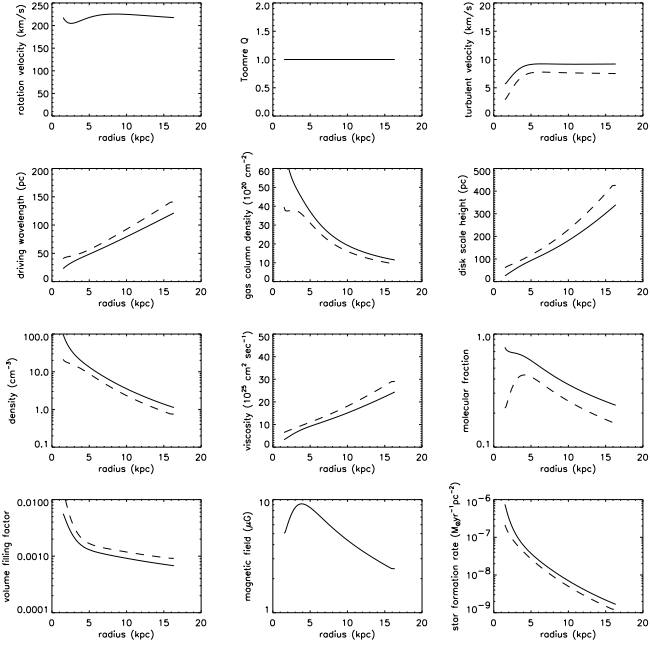}}
        \caption{ \label{fig:numerics_gal_new_sedov}
        SNR model.
        Solid lines: radial profiles of disk parameters for a realistic gravitational
        potential. Dashed lines: disk properties including magnetic and thermal
        pressure.
        }
\end{figure}

In the case of the SNR model
we adopt slightly different parameters in order to match Galactic observations:
$Q=1.0$, $\delta=5$, $\dot{M}=5\,10^{-2}$~M$_{\odot}$\,yr$^{-1}$,
and $\alpha=2\,10^{7}$~yr\,M$_{\odot}$\,pc$^{-3}$.
The resulting radial profiles for the disk parameters are shown in
Fig.~\ref{fig:numerics_gal_new_sedov}.
The total gas mass is $M_{\rm tot}=6.5\,10^{9}$~M$_{\odot}$,
the total molecular mass is  $M_{\rm mol}=2.7\,10^{9}$~M$_{\odot}$, and
the star formation rate is $\dot{M}_{*}=6.5$~M$_{\odot}$\,yr$^{-1}$.
All disk properties have values comparable to those of the SGZ model.
The main differences to the SGZ model are that (i) the volume filling factor
increases significantly with decreasing galactic radius, (ii) the star formation
rate rises more steeply to the galaxy center, and (iii) the molecular fraction
rises less steeply to the galaxy center than for the SGZ model.
As for the SGZ model the gas surface density has approximately a 1/$R$ profile
for $R > 5$~kpc. The star formation rate shows approximately a 1/$R^{2}$ profile.

\subsection{The H{\sc i} to CO surface density relation}

\begin{figure}
        \resizebox{\hsize}{!}{\includegraphics{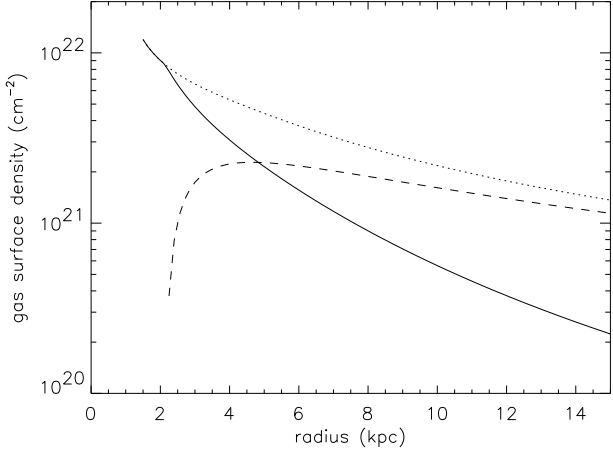}}
        \caption{ \label{fig:profiles}
        Radial profiles of the total (dotted line), molecular (solid line),
        and atomic (dashed line) gas surface density.}
\end{figure}

\begin{figure}
        \resizebox{\hsize}{!}{\includegraphics{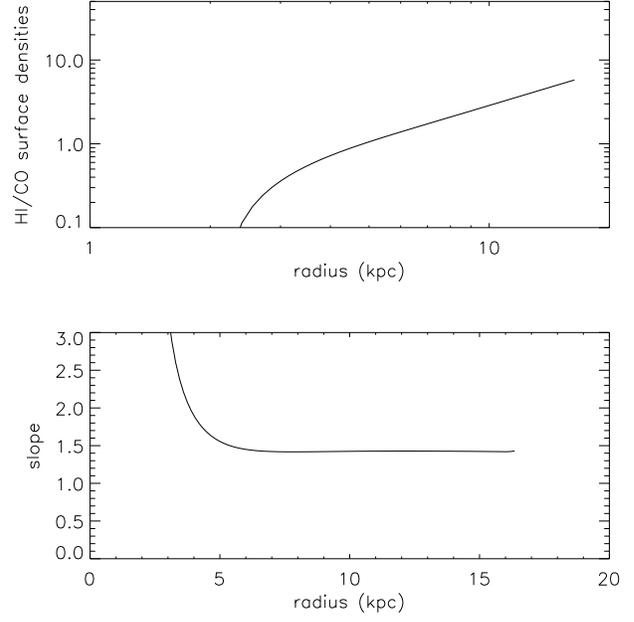}}
        \caption{ \label{fig:HI_CO_profiles}
          The radial dependence of the ratio between the atomic and
          the molecular gas surface density $\Sigma_{\rm HI}/\Sigma_{\rm CO}$
          (upper graph) and its slope (lower graph).
        }
\end{figure}

Fig.~\ref{fig:profiles} shows the radial profiles of the total, molecular,
and atomic gas surface density fro the SGZ model. These profiles well resemble those found
observationally by Wong \& Blitz (2002).
The authors give the following relation between the atomic and the
molecular gas surface density: $\Sigma_{\rm HI}/\Sigma_{\rm CO} \propto R^{1.5}$.
In order to check our model for this relation we show in
Fig.~\ref{fig:HI_CO_profiles} the radial dependence of the ratio
$\Sigma_{\rm HI}/\Sigma_{\rm CO}$ and its slope for the model SGZ.
The curve for the SNR model has approximately the same slope.
For $R > 3$~kpc both models reproduce the observed exponent of 1.5.

\subsection{Magnetic fields and thermal pressure}

For the vertical pressure equilibrium (Eq.~(\ref{eq:pp1}), (\ref{eq:pp2})) we have
neglected the magnetic and thermal pressure: $p_{\rm magn}=B^{2}/(8 \pi)$ and
$p_{\rm th}=n\,k_{\rm B}\,T$, where $B$ is the magnetic field strength, $n$ is the
particle density, $k_{\rm B}$ is the Boltzmann constant and $T$ is the gas temperature.

We have solved the set of equations of Sect.~\ref{sec:realpot} including these
pressure components and setting the turbulent pressure to
$p_{\rm turb}=\frac{1}{2}\,\rho\,v_{\rm turb}^{2}$.
We assume equipartition between the energy density of the turbulent gas
and the magnetic field $\frac{1}{2} \rho v_{\rm turb}^{2} = B^{2}/(8 \pi)$ and
$T$=10\,000~K. The resulting disk properties are plotted as dashed lines in
Fig.~\ref{fig:numerics_gal_new} and \ref{fig:numerics_gal_new_sedov}.

The magnetic field derived from equipartition decreases from $\sim$20~$\mu$G
in the center to $\sim$1~$\mu$G at the edge of the disk. Its value at 8.5~kpc
is $\sim$7~$\mu$G.
Due to the additional pressure terms the turbulent velocity, gas column density,
gas density, molecular fraction, and the
star formation rate decrease, whereas the driving length scale, disk scale height,
and the viscosity increase.
The volume filling factor decreases slightly in the case of the SGZ model and increases
in the case of the SNR model.

The star formation rates are
$\dot{M}_{*} = 3.6$~M$_{\odot}$yr$^{-1}$ for the SGZ model and
$\dot{M}_{*} = 4.2$~M$_{\odot}$yr$^{-1}$ for the SNR model. The radial
profile of the magnetic field then fits nicely that proposed by Strong,
Moskalenko, \& Reimer (2000) in its absolute values. It can be described
by an exponential with a scale length of 8~kpc and a local value
of the magnetic field of $\sim$7~$\mu$G for both models.

\subsection{The star formation law}

\begin{figure}
        \resizebox{\hsize}{!}{\includegraphics{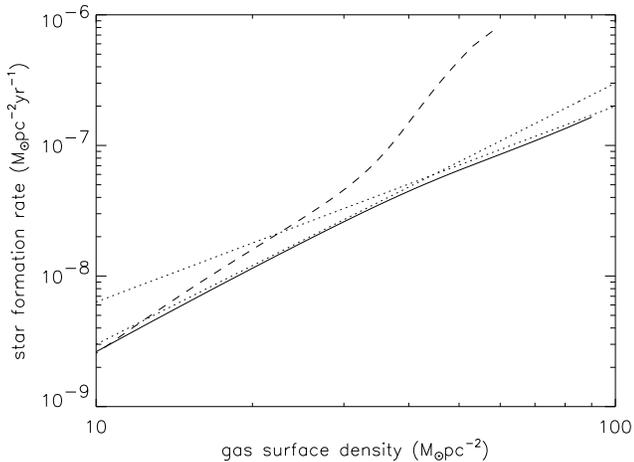}}
        \caption{ \label{fig:schmidt_law}
          The star formation rate $\dot{\Sigma}_{*}$ as a function of the
          total gas surface density $\Sigma$.
          Solid line: SGZ model.
          Dashed line: SNR model
          Dotted lines: $\dot{\Sigma}_{*} \propto \Sigma^{1.5}$ and
          $\dot{\Sigma}_{*} \propto \Sigma^{2}$.
        }
\end{figure}

For the relation between the star formation and the total gas density,
Wong \& Blitz (2002) used a Schmidt law of the form
\begin{equation}
\dot{\Sigma}_{*} \propto \Sigma^{\beta}\ .
\end{equation}
They found $\beta = 1.7 \pm 0.3$ if they use the star formation derived
from H$\alpha$ measurements corrected by an extinction that is proportional to
$\Sigma$. Fig.~\ref{fig:schmidt_law} shows the star formation
rate $\dot{\Sigma}_{*}$ as a function of the total gas surface
density $\Sigma$ for our model together with two curves for
$\beta=1.5$ and $\beta=2$.
The star formation rate of the SGZ model has an exponent $\beta=2$ for
$\Sigma < 50$~M$_{\odot}$pc$^{-2}$ and  $\beta=1.5$ for $\Sigma > 50$~M$_{\odot}$pc$^{-2}$.
The star formation rate of the SNR model has always an exponent greater than 2
and steepens considerably for $\Sigma > 50$~M$_{\odot}$pc$^{-2}$.
Thus, only the SGZ model can reproduce the observational
results of Wong \& Blitz (2002).

\section{Discussion}

The dominating stellar disk (DSD) model applies for gas deficient galaxies
(S0 and cluster galaxies). These are separate cases and will be investigated
in more detail in a separate paper.

The model of a self-gravitating gas disk (SGZ model) and the model where the
SN remnant size is assumed to be the driving length (SNR model) can
be successfully applied to the Galaxy. The input parameters are
$Q \sim $1, $\delta=5$, $\dot{M} \sim 0.05-0.1$~M$_{\odot}$yr$^{-1}$,
$\xi \sim 4.6\,10^{-8}$~(pc/yr)$^{2}$, and
$\alpha \sim 2 - 3\,10^{7}$~yr\,M$_{\odot}$pc$^{-3}$.
The constants of molecule formation $\alpha$ in all models are very close to that
given by Hollenbach \& Tielens (1997).
The fully analytical models (Sect.~\ref{sec:sgz} and \ref{sec:snr}) both
describe the gas properties and the star formation rate of the Galaxy well.
Their radial dependencies are very similar.

The models with a realistic gravitational potential fit observations well and
lead to star formation rates that are consistent with the observed value.
The star formation rate in the inner part of the galactic disk is more realistic
for the SGZ than for the SNR model. Moreover, only the SGZ model
reproduces the observed exponential radial profile of the star formation rate
with the right length scale. Thus, the fit of the radial profile of the star
formation rate favors the SGZ model.

Including a magnetic field in equipartition with the turbulent energy density
leads to a realistic radial profile of the magnetic field in accordance with derived
profiles in the literature.

The main difference between the two models is found in the radial
profile of the volume filling factor of self-gravitating clouds, which
is not well constrained by observations.
Blitz (1993) states that a ``glaring deficiency in
galactic studies of giant molecular clouds is a quantitative study'' of
the radial dependence of their properties.
Thus, both models can be regarded as equally valid.

We conclude that our models correctly reproduce the following
observations:
\begin{itemize}
\item
the integrated total, molecular, and atomic gas mass,
\item
the integrated star formation rate within a factor of 2,
\item
the radial dependence of the ratio between the atomic and molecular
gas surface densities.
\end{itemize}
The exponent for the Schmidt law might be $\sim$20\% too high.

In Sect.~\ref{sec:fullgrav} we show that the viscosity and star
formation prescriptions for the SN driven and the gravity-driven
turbulence (without SN feedback) are equivalent for $Q=1$ and $\delta=1$.
However, the absolute values of the energy flux $\frac{\Delta E}{\Delta t \Delta A}=
-\Sigma \nu \frac{v_{\rm turb}^{2}}{l_{\rm driv}^{2}}$ (Eq.~(\ref{eq:efcturb})) is
different for the two models. In the case of gravity-driven turbulence
the energy flux conservation equation together with the angular momentum
equation and the vertical pressure equilibrium lead to $l_{\rm driv}=H$
independent of a specific viscosity prescription. For the viscosity
$\nu = \gamma v_{\rm turb} l_{\rm driv}$ (Eq.~(\ref{eq:gammavisc})) the gravity-driven
turbulence model reads $\gamma= Re^{-1}$ (Eq.~(\ref{eq:revisc}))
and $H/l_{\rm driv}=1$, whereas the SGZ model
gives $\gamma=1$ and $H/l_{\rm driv}=4.5$. Thus the energy flux due to SN is a factor
of $\sim$200 larger than that due to gravitational instabilities.
This implies that whenever the star formation exceeds 1\% of its
equilibrium value for a given galactic gas disk
(see Sect.~\ref{sec:sgz}), the  energy input due to SN dominates
over that due to gravitational instabilities. Whenever it is smaller
than 1\% of its equilibrium value, gravitational instabilities take over and
drive the ISM turbulence. In this case the mass accretion rate $\dot{M}$
is $\sim$10 times smaller than that of the SN driven turbulent gas disk.

\section{Star formation \label{sec:starformation}}

In this section we study the behaviour of the star formation efficiency
(SFE) defined as the inverse of the star formation time scale:
\begin{equation}
{\rm SFE}=(t_{*})^{-1}=\frac{\dot{\Sigma}_{*}}{\Sigma}\ .
\end{equation}
We are especially interested in the dependence of the SFE
on $Q$.

\subsection{Starbursts ($Q<1$) \label{sec:starbursts}}

There are two possible definitions of a starburst:
\begin{enumerate}
\item
definition based on the absolute value of the star formation rate
measured by the H$\alpha$, radio continuum, or FIR.
In this case, the star formation is enhanced, because of an enhanced
gas surface density, e.g. the Schmidt law ($\dot{\Sigma}_{*} \propto
\Sigma^{\beta}$) is still valid.
\item
definition based on the SFE. In this case the SFE is enhanced (the star formation
time scale is reduced).
\end{enumerate}
Rownd \& Young (1999) studied a sample of starburst galaxies.
Out of 85 galaxies only 24 are starburst galaxies according to definition 2.
Wong \& Blitz (2000) observed one of them, NGC~4736. They found, in agreement
with Rownd \& Young (1999), that the nuclear H$\alpha$ ring does not
follow a Schmidt law, but has a much higher SFE than expected by a
Schmidt law.

Our SGZ model yields
\begin{equation}
{\rm SFE}=\frac{\dot{\Sigma}_{*}}{\Sigma} \propto Q^{-1}
\end{equation}
for the SGZ model and
\begin{equation}
{\rm SFE}=\frac{\dot{\Sigma}_{*}}{\Sigma} \propto Q^{-1.77}
\end{equation}
for the SNR model.
In Sect.~\ref{sec:sgz} we have shown that for the SGZ model
the turbulent velocity dispersion is independent of
$Q$ (Eq.~(\ref{eq:vturbsgz})) and thus $Q$ measures the ratio between the
total enclosed mass and the total gas mass, so that for a given
galactic potential $Q \propto \Sigma^{-1}$. This is also
approximately valid for the SNR model.

If the infall time scale is short enough that a considerable fraction of
gas can be transported to small galactic radii where it is accumulated,
star formation will be enhanced much stronger than
predicted by the Schmidt law in this region. This is one possibility to
create a central starburst. The massive infall needed might be due to
a tidal interaction with another galaxy, an interaction between the
ISM and the intracluster medium in a galaxy cluster, or a bar instability.
Based on Eq.~(\ref{eq:sigmasgz1}) and (\ref{eq:starsgz1}) our model would then predict
\begin{equation}
\dot{\Sigma}_{*} \propto \Sigma^{3}\ .
\end{equation}
However, one has to keep in mind that a $Q<1$ disk is not a stable
configuration and can in principle not be described by an equilibrium disk.
One has to solve
the time dependent equation of transport of angular momentum. Thus, this
estimate must be regarded with caution. Our equilibrium model suggests that
at the beginning of a starburst the turbulent velocity stays constant at
10~km\,s$^{-1}$, whereas the driving length scale is reduced.
This counterbalances the enhanced star formation rate. Evidently, after a
short time the available gas will be consumed and the
SNe might heat the disk and lead rapidly to a $Q>1$ disk.

\subsection{Exhausted star formation ($Q>1$)}

In the opposite case, when gas is taken away from the galaxy by enhanced
star formation or external effects, $Q$ becomes greater than unity.
Then the DSD model applies giving
\begin{equation}
{\rm SFE}=\frac{\dot{\Sigma}_{*}}{\Sigma} \propto Q^{-\frac{2}{5}}
\end{equation}
for the SGZ model.
In this case star formation becomes less efficient, but this effect is
not as pronounced as the increase for $Q<1$ discussed above in
Sect.~\ref{sec:starbursts}.

\section{Summary and Conclusions \label{sec:summary}}

We extend the model of a turbulent cloudy gas disk (Vollmer \& Beckert 2002)
by including energy input through SN explosions.
This is realized (i) by adapting the energy flux conservation equation where
the energy flux transferred by turbulence to smaller length scales
is balanced by the energy input due to SN explosions (SGZ model) or
alternatively (ii) by assuming that the size of a SN remnant equals the
turbulent driving length scale (SNR model).
In the SGZ model the SN energy flux is assumed to be proportional to the local
star formation rate.
The local star formation rate $\dot{\rho}_{*}$ is assumed to be proportional
to the mean density and inversely proportional to the local free fall time
of the clouds. The factor of proportionality is the probability to find
a self-gravitating cloud, i.e. the volume filling factor.
The integration length in $z$ direction is assumed to be the turbulent driving
length scale, i.e. the length scale over which clouds are self-gravitating:
$\dot{\Sigma}_{*}=\dot{\rho}_{*}\,l_{\rm driv}$.

We assume the turbulence to be intermittent due to self-gravity. The turbulence $\nu$
is thus reduced to $\nu = \gamma v_{\rm turb}\,l_{\rm driv}$ with
$\gamma \leq 1$. For $Q \leq 1$ we find $\gamma = 1$.

For model (i) we have calculated two separate cases:
\begin{itemize}
\item
a dominating stellar disk mass (DSD) ($Q > 1$) and
\item
a self-gravitating gas disk in $z$ direction (SGZ) ($Q \sim 1$).
\end{itemize}
For the SGZ model and $\Omega' \sim - \Omega/R$, the radial dependences of
the disk properties (height, turbulent driving length scale, surface density,
density, viscosity, volume filling factor, star formation rate, and molecular
fraction) can be expressed as functions of the mass accretion rate $\dot{M}$,
$Q$, the cloud size divided by the driving length scale $\delta$,
the fraction of SN energy that is transformed into kinetic energy $\xi$,
the constant of molecule formation $\alpha$, and the rotation velocity $\Omega (R)$.
In the case of the DSD model, the dependences on $\Omega (R)$ are the same as for the
SGZ model, whereas those on $\dot{M}$, $Q$, $\alpha$, and $\xi$ are different.
In addition, a factor that depends on the rotation velocity $v_{\rm rot}$ appears.

The SNR model gives radial dependences of the disk properties with
exponents that are close to those of the SGZ model.
Both analytical models (SGZ and SNR) reproduce an almost constant turbulent gas
velocity as observed.

For the SGZ model we conclude that
\begin{enumerate}
\item
the ratio between the turbulent driving length scale $l_{\rm driv}$ and the
disk height $H$ equals the square root of the volume filling factor $\phi_{\rm V}$,
\item
for $Q=1$ the viscosity and star formation prescriptions are equivalent to
those given for the fully gravitational model (Vollmer \& Beckert 2002),
\item
the radial dependences are the same as those for the fully gravitational model.
\end{enumerate}

The SGZ and SNR models are applied to the Galaxy and give a good description of
its gas disk. We derive  $Q \sim 1$ and a mass accretion rate within the disk of
$\dot{M} \sim 0.05 - 0.1$~M$_{\odot}$yr$^{-1}$, which is a factor 5--10 higher than
that of the fully gravitational model.

We include a realistic gravitational potential into the model, which
improves the fit to observations. In this case the SGZ and SNR models
give equally good fits to observations.

The influence of magnetic fields and thermal pressure is investigated.
If both are included in the model, the turbulent velocity decreases with
respect to the case where only the turbulent pressure is taken into account.
An increase of the mass accretion rate by a factor 1.5 makes the turbulent
velocity dispersion increase again to 10~km\,s$^{-1}$.
In this case the radial profile of the magnetic field is close to observations.

The models nicely reproduce the observed radial dependence of the
ratio between the H{\sc i} and molecular surface densities. We derive
a Schmidt law of the form $\dot{\Sigma} \propto \Sigma^{2}$.
A possible explanation for the enhanced star formation efficiency
observed in a few spiral galaxies is given.

We conclude that for $Q=1$ the models for turbulence (i) driven by SN energy
input and (ii) generated by instabilities involving self-gravitation and maintained
by energy input from differential rotation and mass transfer are consistent.
The main difference in the disk properties is that $l_{\rm driv}/H < 1$
for case (i) and $l_{\rm driv}/H = 1$ for case (ii).

Both mechanisms are not exclusive and might possibly coexist in galactic disks.
A galaxy that slowly forms stars will consume its gas and run into the regime
$Q > 1$. Then, the timescale for star formation will increase.
If locally star formation stops entirely, the turbulent
energy will be dissipated within $l_{\rm driv}/v_{\rm turb} \sim 10^{7}$~yr,
the disk shrinks in $z$ direction and ultimately becomes self-gravitating in $z$.
This is the point, where the fully gravitational model of Vollmer \& Beckert
(2002) applies. In this case the driving length scale increases and equals the disk
height. Thus, the energy dissipation rate, which is proportional to $l_{\rm driv}^{-2}$
decreases and the small energy supply due to differential rotation is large
enough to maintain turbulence.

Such a disk has $Q>1$ and a 5 times lower mass accretion rate than a star forming disk.
Once the gas disk is self-gravitating
in $z$ it will form stars again and switch to the SGZ/SNR state.

The occurrence of a bar or an external accretion event
of a companion can lead to a temporal increase of the  mass accretion rate.
If $Q$ becomes smaller than 1 during this event, the star formation timescale
decreases rapidly. The galaxy will then again reach
the regime of $Q \sim 1$. The detailed evolution of such a process can only be
investigated by solving the time dependent equation for angular momentum transport.

\begin{acknowledgements}
We would like to thank W.J. Duschl and A. Shukurov for fruitful discussions and
the anonymous referee for helping us to improve this article significantly.
\end{acknowledgements}

%---------------------------------------------------------------


\begin{thebibliography}{}

\bibitem{a1} Allen, C., \& Santill\'an, A. 1991, RMAA, 22, 255

\bibitem{a1a} Avillez M.A. \& Mac Low M.-M. 2001, ApJ, 551, L57

\bibitem{a2} Binney J. \& Tremaine S. 1987, in Galactic Dynamics, Princeton University Press

\bibitem{a2a} Blitz L. 1993, in: Protostars and planets III (A93-42937 17-90), p. 125-161.

\bibitem{a2b} Boulares A. \& Cox D.P. 1990, ApJ, 365, 544

\bibitem{a2c} Dorfi E.A. 1993, in: Galactic High-Energy Astrophysics. High-Accuracy Timing and Positional Astronomy. Lectures Held at the Astrophysics School IV Organized by the European Astrophysics Doctoral Network (EADN) in Graz, Austria, 19-31 August 1991, Edited by Jan van Paradijs and Hans M. Maitzen. Springer-Verlag Berlin Heidelberg New York. Volume 418, p.43

\bibitem{a3} Elmegreen B.G. \& Falgarone E. 1996, ApJ, 471, 816

\bibitem{a4} Frisch U. 1995, Turbulence -- The Legacy of A.N. Kolmogorov, Cambridge University press

\bibitem{a5} Gazol-Patino A. \& Passot T. 1999, ApJ, 518, 748

\bibitem{a6} Hollenbach D.J. \& Tielens A.G.G.M. 1997, ARA\&A, 35, 179

\bibitem{a7} Korpi M.J., Brandenburg A., Shukurov A., Tuominen I., \& Nordlund A. 1999, ApJ, 514, L99

\bibitem{a8} Kulkarni S.R. \& Heiles C. 1988, in: Galactic and Extragalactic Radio Asrtonomy, ed. G.L. Verschurr \& K.I. Kellermann (Springer, Berlin), 95

\bibitem{a9} Landau L.D. \& Lifschitz E.M. 1959, Fluid Mechanics

\bibitem{a10} McKee C.F. 1995 in: The Physics of the Interstellat Medium and the
Intergalactic Medium, ASP Conf. Ser. 80, ed. A. Ferrara, C. Heiles, C. McKee, \& P. Shapiro (ASP, San Fransisco), 292

\bibitem{a11} Paczy\'{n}ski B. 1978, Acta Astron., 28, 91

\bibitem{a12} Prantzos N. \& Aubert O. 1995, A\&A, 302, 69

\bibitem{a12a} Rosen A. \& Bregman J.N. 1995, ApJ, 440, 634

\bibitem{a13} Rownd B.K. \& Young J.S. 1999, AJ, 118 670

\bibitem{a13a} Ruzmaikin A.A., Shukurov A.M., \& Sokoloff D.D. 1988, Magnetic Fields of Galaxies, Astrophysics and space science library, Vol. 133, Kluwer Academic Publishers

\bibitem{a14} Scalo J. 1985, in Protostars and Planets II, ed. D.C. Black \& M.S. Matthews (Tucson: Univ. Arizona), 201

\bibitem{a15} Spitzer L. 1990, ARA\&A, 28, 71

\bibitem{a15a} Strong A.W., Moskalenko I.G., \& Reimer O. 2000, ApJ, 537, 763

\bibitem{a16} Tammann G.A., L\"{o}ffler W., \& Schr\"{o}der A. 1994, ApJS, 92, 487

\bibitem{a17} Thornton, K., Gaudlitz, M., Janka, H.-Th., \& Steinmetz, M. 1998, ApJ, 500, 95

\bibitem{a18} Toomre A. 1964, ApJ, 139, 1217

\bibitem{a19} Vollmer B. \& Beckert T. 2002, A\&A, 382, 872

\bibitem{a20} Wada K. \& Norman C.A. 2001, ApJ, 547, 172

\bibitem{a20a} Wada K., Meurer G., \& Norman C.A. 2002, ApJ, 577, 197

\bibitem{a21} Wong T. \& Blitz L. 2000, ApJ, 540, 771

\bibitem{a22} Wong T. \& Blitz L. 2002, ApJ, 569, 157

\end{thebibliography}
\end{document}